# Insight into the molecular dynamics of barocaloric molecular crystals using quasi-elastic neutron scattering

Frederic Rendell-Bhatti[1,*], David Boldrin[1], Markus Appel[2] and Donald A. MacLaren[1]

[1]SUPA, School of Physics and Astronomy, University of Glasgow, Glasgow, UK

[2]Institut Laue-Langevin (ILL), 71 Av. Des Martyrs, 38000 Grenoble, France

*Email Correspondence: fred.rendell@glasgow.ac.uk

**ORCIDs**
F.R.B.: https://orcid.org/0000-0002-3470-786X, D.B.: https://orcid.org/0000-0003-3833-8341, D.A.M.: https://orcid.org/0000-0003-0641-686X, M.A.: https://orcid.org/0000-0001-6069-7573

**Key words:** Caloric, plastic crystal, solid-solid phase transition, QENS, rotational dynamics

# Abstract

Barocaloric (BC) molecular crystals exhibit order-disorder phase transitions that offer a promising basis for solid-state heating and cooling. However, rational design of these next-generation materials requires microscopic understanding of the molecular dynamics that drive the transitional entropy change. Quasi-elastic neutron scattering (QENS) techniques including inelastic fixed-window scans (IFWSs) enable the direct probing of such molecular motions throughout thermodynamic phase transitions. Here we present a QENS-IFWS global fitting methodology for systematically investigating the molecular dynamics underpinning BC performance in molecular crystals. We demonstrate this methodology on neopentyl glycol (NPG) and three derivatives spanning chemically distinct routes to dynamical perturbation. By fixing attempt frequencies across samples, we observe measurable differences in the relative activation energies and mobile fractions of hydroxymethyl and full-molecular rotations. The largest observed effect is on the hydroxymethyl rotation activation energies, which is closely related to hydrogen bond network stability, ranging from -3.6% to +11.1% in the derivatives compared to pure NPG. The capability of this global fitting strategy to extract subtle variations in molecular rotations highlights QENS as a versatile design tool for guiding BC molecular crystal design.

# 1 Introduction

The transition to low-carbon heating and cooling technologies requires the development of refrigerant materials with low environmental impact and competitive thermodynamic performance. Conventional vapour-compression refrigerants, while widely deployed by the heating and cooling industry, are gases that can be flammable or toxic and, crucially, have appreciable global warming potential (GWP) when they inevitably leak into to the atmosphere [1,2]. Solid-state caloric materials, which exhibit reversible entropy changes in response to applied external fields [3], present a promising alternative pathway to solid-state heating and cooling technologies [4,5], avoiding problematic refrigerant leakage. The barocaloric (BC) effect [6], which is the entropy change associated with the application of hydrostatic pressure, is attractive due to its applicability to a wide range of materials with varying use-cases [7–12], the maturity of pressure-driven engineering [12], and the potential for BC technologies to exceed the efficiency of current vapour-compression systems [13,14].

Within the BC materials landscape, molecular plastic crystals have emerged as a leading class of solid-state refrigerants [15–25]. These materials exhibit so called 'colossal' entropy changes (>100 J kg$^{-1}$ K$^{-1}$) associated with order-disorder solid-solid phase transitions in which the high-temperature plastic crystal (PC) phase involves substantial rotational disorder of the constituent molecules. Neopentyl glycol (NPG, $(CH_3)_2C(CH_2OH)_2$) is a prototypical example, exhibiting a solid-state transition from a monoclinic ordered crystal (OC) phase to a face centred cubic PC phase near 314 K [26,27]. NPG was shown to exhibit one of the largest BC entropy changes reported in this materials class [15,16], demonstrating a real opportunity for BC materials to replace gaseous refrigerants. However, sizeable thermal hysteresis in the PC to OC supercooled transition limits the theoretical efficiency of these materials during a refrigeration cycle [28,29]. The origins of this hysteresis remain poorly understood, but reduction of hysteresis whilst preserving the entropy change is a research priority. Rational design of next-generation BC molecular crystals, including doped, composite and functionalised variants, requires direct mechanistic insight into the molecular motions activated at the transition.

Several experimental and computational techniques offer microscopic information on molecular dynamics (MD), including solid-state nuclear magnetic resonance (NMR) [30,31] and dielectric and infrared spectroscopy [32,33]. MD simulations provide complementary atomistic information but are typically limited in their ability to span the full range of timescales relevant to the transition [27,34–37]. Quasi-elastic neutron scattering (QENS) provides a direct experimental probe of incoherent hydrogen motions on picosecond–nanosecond timescales and is particularly well-suited to hydrogenous molecular crystals because hydrogen dominates the incoherent QENS signal [38,39]. Previous QENS studies have characterised specific rotational modes in BC molecular crystals [16,35,40,41], including NPG and related molecules, but the systematic application of QENS to chemically perturbed variants for materials-design purposes has been limited.

In this work, we demonstrate a comparative QENS and inelastic fixed-window scan (IFWS) global fitting methodology for the molecular dynamics of BC molecular crystals and apply it to pure NPG and three chemically distinct derivatives: NPG with pentaerythritol (NPG+PE), a ternary solid solution of NPG with pentaglycerine and pentaerythritol (NPG+PG+PE), and an NPG composite with carboxylated detonation nanodiamond (NPG+DND). The first two materials have been previously shown to exhibit reduced thermal hysteresis compared to pure NPG [41,42], whilst the latter contains dispersed nanodiamonds that have been shown to modify the hydrogen bonding in water [43] and is expected to similarly affect hydrogen bonding in NPG. The hydrogen bond networks in NPG, NPG+PE and NPG+DND are determined by the monoclinic crystal structure of NPG, where hydrogen bond ladders repeat periodically along the *a* axis [27]. By contrast, the hydrogen bond network in NPG+PG+PE is determined by its body centred tetragonal crystal structure [40], where the intermolecular hydrogen bonds form in the *ab* plane, with weak Van der Waals interactions between layers. In both cases, the existence of the hydrogen bond networks coincides with the ability of the hydroxymethyl groups in these molecules to rotate and thus indicates when the networks form and break during the solid-solid phase transitions.

In these systems, the QENS spectra are dominated by three distinct rotational modes of the NPG molecule: methyl rotation, hydroxymethyl rotation and full molecular reorientation [40]. By globally fitting the QENS spectra and the IFWS scans with shared Arrhenius parameters, and by constraining the attempt frequencies to their pure-NPG values across samples, we resolve these three rotational modes (methyl rotation, hydroxymethyl rotation, and whole-molecule reorientation) and quantify their relative activation energies and elastic incoherent structure factor (EISF)-derived mobile fractions across all four samples. The global fitting methodology has been outlined previously by us [40,41], and others [44], but here we seek to impose additional constraints on the large parameter space of QENS-IFWS global fitting to enable the comparison of subtle molecular dynamics differences between related materials. We show that the samples follow mechanistically distinct routes for perturbing the molecular dynamics associated with the hydrogen bond network, demonstrating QENS as a versatile design tool for guiding the development of next-generation BC molecular crystals.

# 2 Materials and Methods

## 2.1 Sample Preparation

Four samples were prepared for this study: pure neopentyl glycol (NPG); NPG doped with pentaerythritol (PE), with molar ratio $NPG_{0.97}PE_{0.03}$; NPG doped with 5% by mass carboxylated detonation nanodiamond (DND); and a ternary solid solution of NPG with pentaglycerine (PG) and PE, with molar ratio $NPG_{0.60}PG_{0.38}PE_{0.02}$. NPG (99% purity), monodispersed carboxylated DND (5 nm), PE (99% purity) and PG (99% purity) were obtained from Sigma Aldrich and used as received. NPG+PE was prepared by co-dissolution of NPG with PE in ethanol, followed by recrystallisation under ambient conditions. NPG+DND was prepared by dissolving pure NPG in a carboxylated nanodiamond water dispersion and left to evaporate under ambient conditions to produce a composite. The ternary NPG+PG+PE solid solution was prepared by co-dissolution of NPG, PG, and PE, followed by recrystallisation under ambient conditions to produce a single-phase solid solution. PG and PE are both neopentyl small-molecules and are related to NPG by the number of hydroxymethyl groups, where NPG has two, PG has three and PE has four, and as such these molecules display very similar QENS signals.

## 2.2 QENS Measurements

QENS and FWS measurements were performed on the IN16B backscattering spectrometer at the Institut Laue-Langevin (ILL), Grenoble, France. IN16B was used in standard configuration with a strained Si111 Doppler monochromator and analysers [45], providing an instrumental energy resolution of 0.75 μeV (FWHM linewidth) and a dynamic range of ±28 μeV, accessing localised molecular motions with characteristic times approximately between 10 ps and 1 ns. The full accessible momentum transfer range was $0.19 \leq Q \leq 1.89$ Å$^{-1}$.

Powder samples of masses between 200 to 500 mg were lightly pressed into thin (~ 1mm) films between aluminium foil at 330 K and subsequently wrapped around the cylindrical aluminium sample cans. Vanadium, empty cell and resolution measurements were obtained with the same sample can geometry. Elastic and inelastic intensities were sampled for each material using elastic fixed-window scans (EFWS) at $E$ = 0 μeV and IFWS at $E$ = ±3 μeV, recorded continuously on heating and cooling at a rate of ~ 1 K min$^{-1}$. Full QENS spectra were collected at a series of fixed temperatures spanning the OC and PC phases. For NPG+PG+PE, an additional QENS spectrum was collected within the hysteresis region of this sample. The instrumental resolution function for each sample was determined by a separate QENS measurement at a base temperature of ~2 K, where all rotational modes are frozen on the IN16B timescale. Standard data reduction (detector normalisation, background subtraction, and energy calibration) was carried out using standard procedures in the Mantid software [46].

## 2.3 QENS Analysis

The QENS signal from a powder sample is the orientationally averaged dynamic structure factor, $S(Q,E)$, giving the scattering intensity as a function of momentum transfer, $Q$, and energy transfer, $E$:

$$S(Q,E) = [A_0(Q)\delta(E) + \Sigma_i A_i(Q)L_i(E)] \otimes R(Q,E) + B(Q). \qquad (1)$$

For a system undergoing localised rotational stochastic motions, $S(Q,E)$ decomposes into an elastic delta-function, $\delta(E)$, with an amplitude $A_0(Q)$ and one or more quasi-elastic Lorentzian components $L_i(E)$, each associated with a distinct rotational mode, $i$, with a temperature-dependent FWHM linewidth, $\Gamma_i(T)$, and amplitude, $A_i(Q)$. These sample signals are convolved with the instrumental resolution, $R(Q,E)$, and $B(Q)$ is a flat background. The temperature dependence of each Lorentzian linewidth is assumed to follow an Arrhenius law:

$$\Gamma_i(T) = \Gamma_{0,i} \cdot \mathrm{e}^{\frac{-E_{a,i}}{k_B T}}, \qquad (2)$$

where the prefactor, $\Gamma_{0,i}$, is related to the attempt frequency through $v_{0,i} = \frac{\Gamma_{0,i}}{h}$ and $E_{a,i}$ is the activation energy, all for the $i^{th}$ mode. The $Q$-dependence of the intensity ratio between the elastic peak and the total spectrum is known as the elastic incoherent structure factor (EISF) and for a single mode takes the form:

$$\mathrm{EISF}^{\mathrm{exp}}(Q) = \frac{A_0(Q)}{A_0(Q)+A_1(Q)}. \qquad (3)$$

**Equation 3** encodes the geometry of the corresponding motion described by the quasi-elastic signal and can be identified via the form of the geometric model, $\mathrm{EISF}^{\mathrm{geom}}(Q)$, used to fit the $\mathrm{EISF}^{\mathrm{exp}}(Q)$ data. A fractional parameter, $f$, is often included to account for the possibility that only a fraction of scatterers participate in the respective process [40,41,47–49]:

$$\mathrm{EISF}^{\mathrm{geom,observed}}(Q) = (1-f) + f \cdot \mathrm{EISF}^{\mathrm{geom}}(Q). \qquad (4)$$

Inelastic fixed-window scans (IFWS) complement full-spectrum QENS by measuring the scattering intensity at a fixed non-zero energy transfer as a function of temperature [45]. At a chosen offset energy, $\pm E_{\mathrm{off}}$, the IFWS signal of a given Lorentzian mode peaks when $\Gamma_i(T) \approx 2E_{\mathrm{off}}$, producing a characteristic temperature signature for each mode. Crucially, because the same Arrhenius parameters describe the temperature dependence of $\Gamma_i(T)$ in both the QENS spectra and the IFWS scans, the two data sets can be co-fitted with shared parameters, which can be a powerful internal consistency check that constrains the dynamical model more robustly than either measurement alone. Full details of the global fitting are outlined in **Supplementary Note S1** and are summarised below.

QENS spectra and IFWS scans were analysed within the Mantid software using custom global-fit functions, which are provided as **Supplementary Information Files**. The model for the QENS dynamic structure factor $S(Q,E)$ (**Eq. 1**) consisted of a delta-function elastic peak, a number of Lorentzians representing the rotational modes resolved at a given temperature convolved with the experimentally determined instrumental resolution function, and a flat background. The flat background term was included to account for inelastic signals outside the dynamic range of IN16B (see **Supplementary Note S1.5**). The IFWS signal at $E_{\mathrm{off}}$ = ±3 μeV was modelled by evaluating the same $S(Q,E)$ at the IFWS offset energy window. The Arrhenius parameters ($\Gamma_{0,\mathrm{i}}$ and $E_{\mathrm{a,i}}$ in **Eq. 2**) for each mode were shared between the QENS and IFWS components of the global fit, so that a single set of Arrhenius parameters per mode simultaneously described both the energy-resolved QENS measurements and the temperature-resolved IFWS scans (see **Supplementary Note S1.1-S1.2**). Data were $Q$-resolved into individual bins ($\Delta Q \approx 0.1$ Å$^{-1}$) and fitted across a $Q$-range of 0.29 Å$^{-1}$ < $Q$ < 1.84 Å$^{-1}$ for the global fits. EISF analysis was performed over the same $Q$-range, where the OC phase EISF was calculated according to **Eq. 3** and **Eq. 4**, and the PC phase EISFs took a modified form to account for multiple observed dynamical modes. Further details of the EISF fitting procedure, the explicit functional forms of the IFWS and QENS model components, and the strategy for fixing the PC phase attempt frequencies are given in the **Results & Discussion** and expanded upon in **Supplementary Note S1**.

# 3 Results and Discussion

## 3.1 QENS and IFWS Global Fitting

**Figure 1a-d** presents a comparative overview of the QENS and FWS data for NPG, NPG+PE, NPG+DND and NPG+PG+PE, with complete data found in **Supplementary Figs. S1-S2**. The plots shown in **Fig. 1a-d** show the $Q$-summed (0.59 Å$^{-1}$ ≤ $Q$ ≤ 1.84 Å$^{-1}$) raw experimental data; EFWS, IFWS, and selected OC and PC QENS spectra, respectively, for all four samples on normalised intensity scales. **Figure 1a,b** shows FWS measurements, providing scattering intensity as a function of temperature, $S(T)$, between

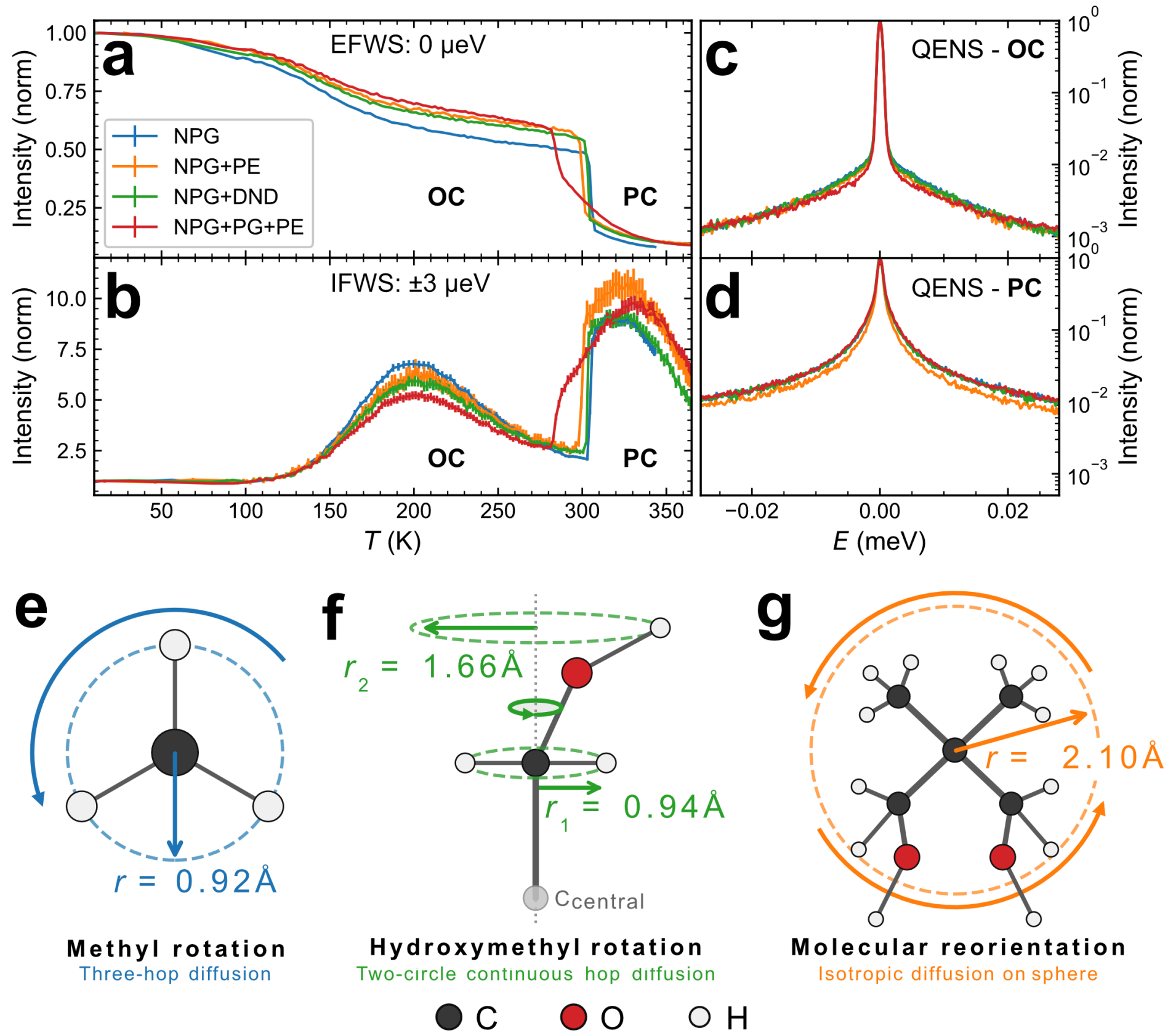

**Figure 1.** Overview of neutron scattering measurements and the geometric models used in their interpretation. (a–d) raw experimental data for 0.59 Å$^{-1}$ ≤ $Q$ ≤ 1.84 Å$^{-1}$, showing all samples on normalised intensity scales. (a) Elastic and (b) inelastic fixed-window scans on cooling, intensity normalised at low-$T$. Representative QENS spectra in the (c) OC phase and (d) PC phase. Sample colours are shared across (a–d). (e–g) Schematics of the three rotational modes resolved by QENS and the radii used to model their rotational geometry. (e) Methyl rotation, modelled as three-fold jump diffusion on a circle of radius $r$ = 0.92 Å. (f) Hydroxymethyl rotation, modelled as two-circle continuous diffusion about the C–C bond axis (vertical dotted line): the two $CH_2$ hydrogens trace an inner circle of radius $r_1$ = 0.94 Å and the OH hydrogen traces an outer circle of radius $r_2$ = 1.66 Å. (g) Whole-molecule reorientation, modelled as isotropic continuous diffusion on a sphere of effective radius $r$ = 2.10 Å, approximating isotropic rotation about the molecule centre of mass.

~10 K and ~360 K at fixed energies, for elastic ($E$ = 0 meV) and inelastic (E = ±3 meV) measurements, respectively. In contrast, **Fig. 1c,d** provides full energy spectrum QENS measurements, $S(E)$, at discrete temperatures below and above the phase transition, respectively.

Inspecting the EFWS cooling data in **Fig. 1a**, the almost linear intensity drop of the EFWS intensity at very low temperatures ($T$ < 100K) is attributed to Debye-Waller effects, before the onset of quasi-elastic scattering effects in the dynamic range of IN16B. This region displays very small variations between samples, suggesting there is minimal differences associated with atomic vibrations. Both the EFWS (**Fig. 1a**) and IFWS (**Fig. 1b**) clearly show the PC-OC phase transition as a sharp drop(increase) in

elastic(inelastic) intensity around 300 K, with the precise transition temperature and width varying between samples. This sharp change in FWS intensity corresponds to the appearance and disappearance of the PC modes (hydroxymethyl rotation and molecular reorientation), whose existence is correlated with the hydrogen bond network [16,40,41]. The EFWS data around the hysteresis region on heating and cooling is shown in **Supplementary Fig. S3** and demonstrates how FWS measurements can directly probe the molecular dynamics onset and thermal hysteresis temperatures in BC molecular crystals. The IFWS in **Fig.1b** exhibits two characteristic intensity features for each sample: a broad peak near 200 K, arising from the methyl rotation crossing through the ±3 μeV window, and a sharp increase at the OC-PC transition, signalling the onset of the hydroxymethyl rotation and molecular reorientation modes of the PC phase. The QENS spectra in **Fig.1c,d** confirm this picture, with substantially broader QENS wings in the PC phase reflecting the accelerated rotational dynamics. Given the fact that all the measured samples are mostly composed of NPG, it is expected that the QENS signals shown in **Fig.1a-d** arise from the same subset of dynamical modes.

**Figure 1e–g** summarises the three expected rotational modes accessed by QENS on IN16B in NPG and the geometric radii used to model their EISFs. The methyl rotation (**Fig. 1e**) operates throughout both the OC and PC phases but is directly resolved as a Lorentzian only in the OC phase; in the PC phase its linewidth exceeds the IN16B dynamic range [16,40], and the mode instead contributes to the flat inelastic background (see **Supplementary Information Note S1.5**). Both the hydroxymethyl rotation (**Fig. 1f**) and molecular reorientation (**Fig. 1g**) are liberated or frozen at the phase transition of the material on heating and cooling, respectively, and are responsible for the sharp change in quasi-elastic signal between the OC and PC phases shown in **Fig. 1a-d**. These geometric models are the same as those used to study QENS of NPG and NPG+PG+PE previously, except for the model describing hydroxymethyl rotation, shown in **Fig. 1f**. Previously, this mode was modelled as continuous jump diffusion on a circle of all three hydroxymethyl hydrogens, using a single mean rotation radius [40,41]. Here, we develop this model further by accounting for the fact that the hydroxymethyl group hosts two distinct hydrogen environments; the -$CH_2$ and -OH hydrogens, with different rotational radii ($r_1$ and $r_2$ in **Fig. 1f**). Further details on the geometric models used for EISF fitting can be found in **Supplementary Note S1.3**. It is important to note that the decomposition of the PC-phase quasi-elastic signals into independent hydroxymethyl rotation and molecular reorientation Lorentzians is an approximation. The -$CH_2$ and -OH hydrogens will participate simultaneously in both motions, such that their true scattering law is a convolution of the two contributions rather than a sum. This treatment is valid in the well-separated-timescale limit, which holds here as the two modes differ by roughly an order of magnitude in characteristic time [40].

The complementary information encoded in the QENS spectra and the IFWS scans is illustrated in **Fig. 2**, showing representative samples NPG (**Fig. 2a**) and NPG+PG+PE (**Fig. 2b**) summed over $0.59\ Å^{-1} \leq Q \leq 1.84\ Å^{-1}$. The global fitting of QENS and IFWS datasets exploits the fact that both methods are probing the total quasi-elastic scattering of the sample, $S(Q,E,T)$, where QENS operates at fixed temperatures; $S_{QENS}(Q,E)$, and IFWS operates at fixed energy offsets; $S_{IFWS}(Q,T)$. As such, when combined, they offer complementary information when their observable parameters are linked through the Arrhenius relationship (given in **Eq. 2**) during global fitting. This methodological strategy is expanded upon further in the **Methods** and **Supplementary Note S1.1**.

Inspecting **Fig. 2a,b**, six QENS spectra spanning the OC and PC phases are plotted in constant-$T$ planes, with the simultaneously measured IFWS at $E$ = +3 μeV plotted in the constant-$E$ plane. Black markers indicate the ($T$, +3 μeV) coordinates at which the QENS and IFWS measurements sample the same physical observable, demonstrating the self-consistency of the two data types. The sharp rise in IFWS intensity at the OC-PC transition shares its origin with the increase in QENS intensity at +3 μeV, motivating the co-analysis of the two data sets. For NPG (**Fig. 2a**) the transition is sharp, with a clean change in mode populations between adjacent measurement temperatures. For NPG+PG+PE (**Fig. 2b**) the transition is significantly broader, with the 293 K QENS measurement (indicated by the black arrow) sitting inside the hysteresis region (see **Supplementary Fig. S3d**). This QENS scan is excluded from the global Arrhenius fit and is instead analysed separately (**Supplementary Fig. S10**) to characterise the intermediate-state dynamics, where it is found that both OC and PC modes coexist at this temperature.

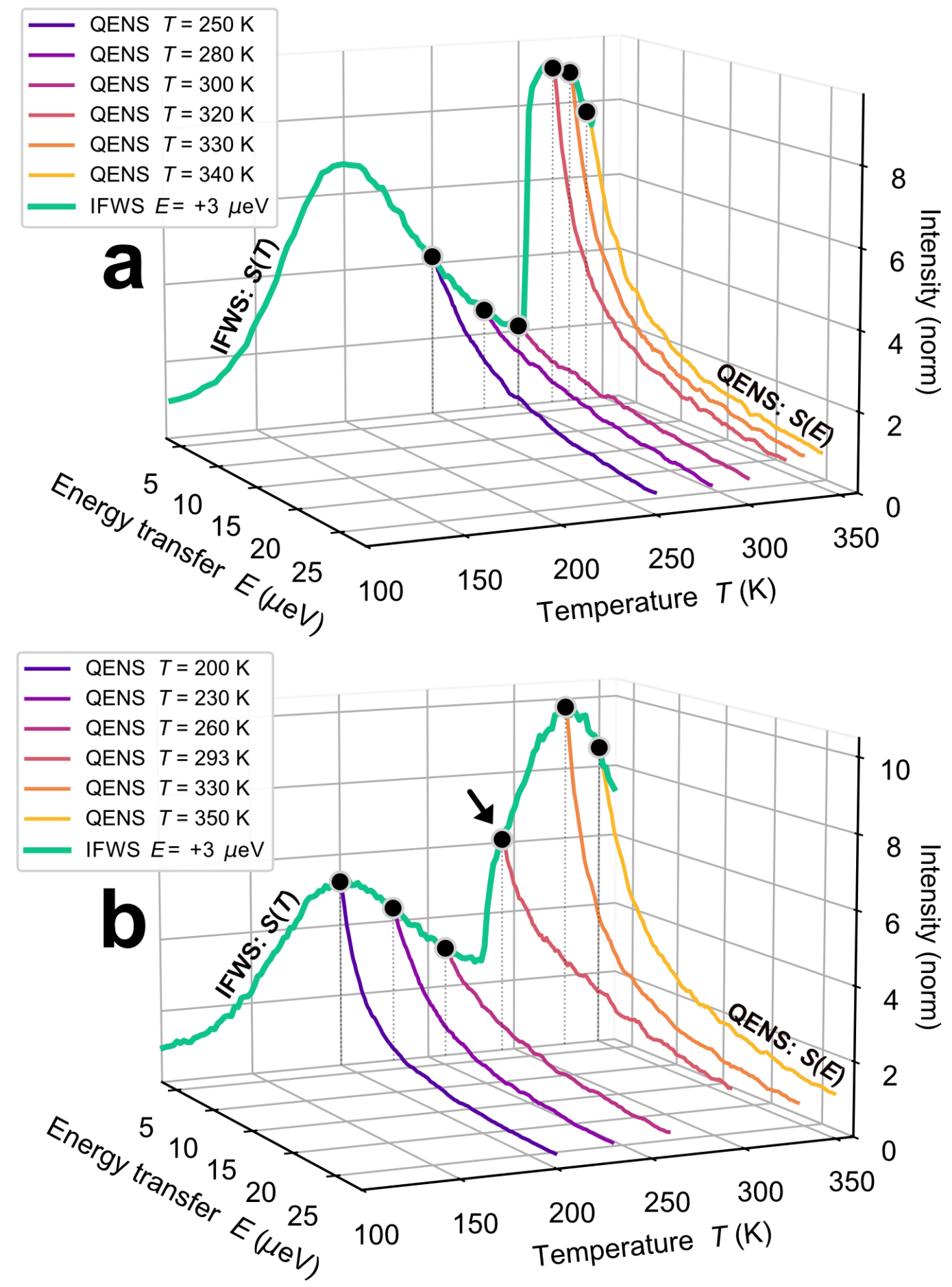

**Figure 2.** Illustration of the complementary information encoded in QENS and IFWS measurements of (a) NPG and (b) NPG+PG+PE for 0.59 $Å^{-1} \leq Q \leq 1.84$ $Å^{-1}$. QENS spectra (purple to yellow) are plotted in constant-$T$ planes at six temperatures spanning the OC phase and the PC phases. The IFWS scans (green) are plotted in the constant $E$ = +3 μeV plane and are recorded continuously through the transition on cooling. Black markers indicate the ($T$, +3 μeV) coordinates at which QENS and IFWS measurements sample the same physical observable. (a) The sharp rise in IFWS intensity across the transition around 300 K reflects the liberation of the hydroxymethyl and molecular reorientation modes. (b) The phase transition is much less sharp than in NPG, reflecting the broad phase co-existence region of NPG+PG+PE. The QENS spectra at 293 K (indicated by the black arrow) is located within the hysteresis region and displays intermediate mode populations.

**Figure 3** shows representative global QENS+IFWS fits for NPG+PG+PE; analogous fits for all samples are provided in the electronic supplementary material (**Supplementary Figs. S4–S15**). In the OC phase, a single Lorentzian representing methyl rotation (**Fig. 1e**) suffices to describe both the QENS spectrum (**Fig. 3a**) and the IFWS scan (**Fig. 3c**), where the characteristic IFWS peak near 200 K arises from this methyl rotation. In the PC phase (**Fig. 3b,d**), two Lorentzians (orange and green lines) are required: a narrow component representing whole-molecule reorientation (**Fig. 1g**) and a broader component representing hydroxymethyl rotation (**Fig. 1f**). The IFWS intensity in this temperature range shows the simultaneous contributions of both modes, with the hydroxymethyl peak occurring near 315 K and the molecular reorientation peak located at $T > 340$ K. The total fit (red line) in each panel is the result of the same $\Gamma_0$ and $E_a$ $Q$-independent parameters for the QENS and IFWS components of each mode, demonstrating the internal consistency of the global fit and motivating the comparative methodology discussed in the remainder of this section.

**Table 1** provides the activation energies and Arrhenius prefactors extracted from the global fits for all samples and modes. Because $\Gamma_0$ and $E_a$ are strongly correlated in an Arrhenius fit over a limited temperature range, a small change in $\Gamma_0$ can be compensated by a small change in $E_a$ to leave the resulting $\Gamma(T)$ essentially unchanged. Therefore, the absolute $E_a$ values of the PC modes carry a larger systematic uncertainty than the statistical uncertainties quoted in **Table 1**, since their absolute value depends on the precise choice of $\Gamma_0$ for each mode. To make the cross-sample comparison statistically meaningful, the mode attempt frequencies were fixed at the best-fit values (with limits between ~$10^{10}$ Hz and ~$10^{14}$ Hz) obtained from pure NPG, also presented in **Table 1**. Systematic uncertainties and are discussed further in **Supplementary Note S1.2**. With this constraint, only the activation energies vary between samples, and the resulting relative variations are statistically precise, giving uncertainties of only ±0.1 to ±0.7 meV across the modes and samples. Note that the systematic uncertainty associated with fixing $\Gamma_0$ is common to all samples and cancels in the cross-sample differences, leaving an internally consistent comparison.

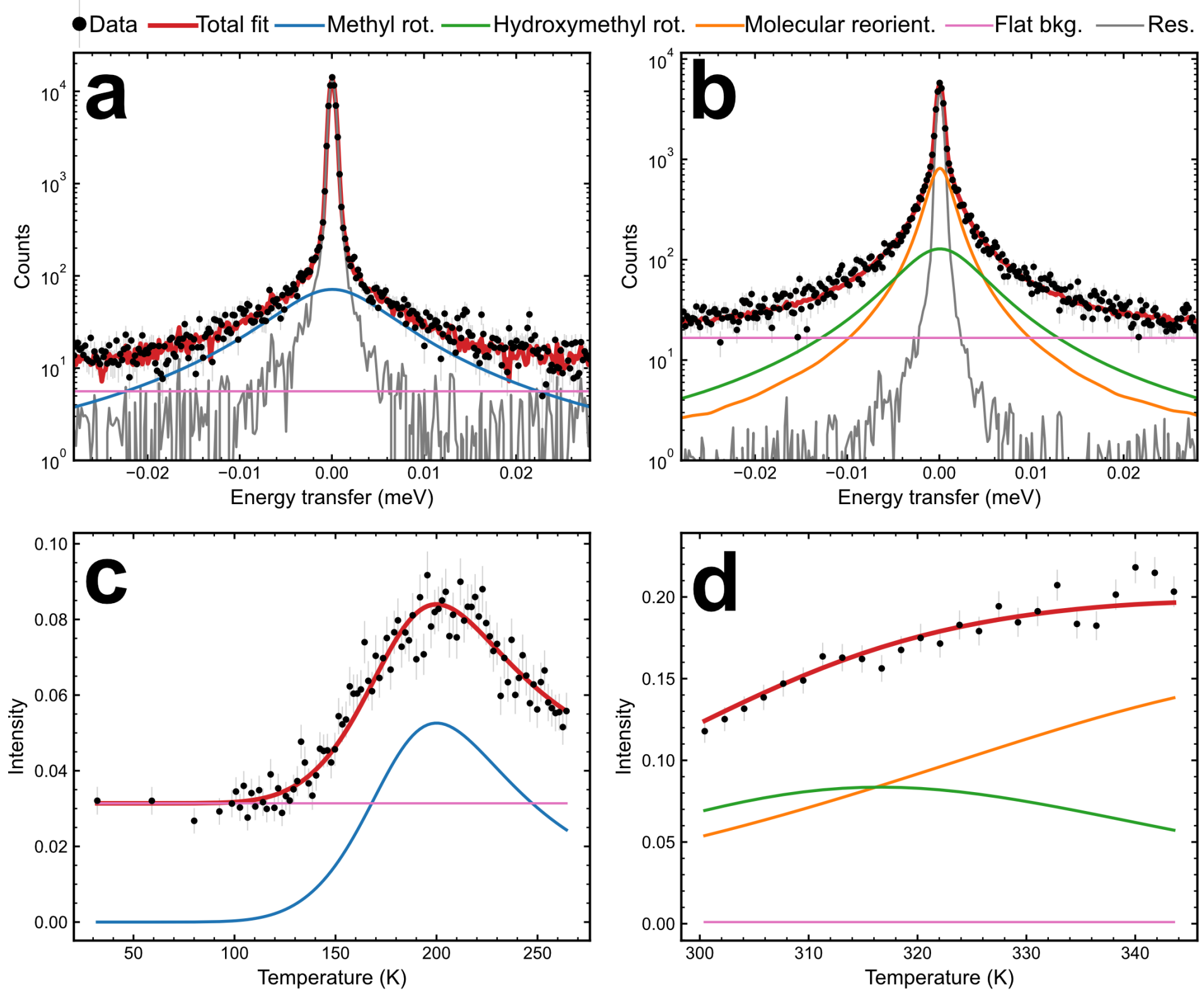

**Figure 3.** Representative ($Q$ = 0.71 Å$^{-1}$) global QENS-IFWS fits for NPG+PG+PE. (a) OC phase QENS spectrum at $T$ = 230 K, fitted with a single Lorentzian (blue) representing methyl rotation, convolved with the instrumental resolution function (grey), plus a flat background (pink). (b) PC phase QENS spectrum at $T$ = 330 K, fitted with two Lorentzians representing molecular reorientation (orange) and hydroxymethyl rotation (green), in addition to the resolution function and flat background. (c) IFWS in the OC phase (cooling 270 K to 50 K), where the temperature dependence is dominated by the methyl rotation contribution (blue). (d) IFWS in the PC phase (cooling 345 K to 300 K), showing the simultaneous contributions of molecular reorientation (orange) and hydroxymethyl rotation (green).

The Arrhenius prefactors in **Table 1** when converted to frequency units are approximately $\upsilon_{0,1} \approx 2.3 \times 10^{11}\text{Hz}$ for methyl rotation, $\upsilon_{0,2} \approx 9.5 \times 10^{13}\text{Hz}$ for hydroxymethyl rotation and $\upsilon_{0,3} \approx 2.5 \times 10^{12}\text{Hz}$ for molecular reorientation, which are consistent with rotational attempt frequencies [50]. Inspecting the spread in activation energies for each mode in turn across the samples; methyl rotation ranges between $98.4\text{ meV} \leq E_{a,1} \leq 100.4\text{ meV}$, hydroxymethyl rotation ranges between $278.8\text{ meV} \leq E_{a,2} \leq 321.5\text{ meV}$ and molecular reorientation ranges between $239.1\text{ meV} \leq E_{a,3} \leq 255.1\text{ meV}$. The fact that the hydroxymethyl rotation has a higher activation energy compared to molecular reorientation is surprising, since the hydroxymethyl rotation must be liberated before the whole molecule is able to rotate. This trend has been observed previously both experimentally [40] and theoretically [27] for pure NPG, and we have previously explained it by the fact that the attempt frequency of the hydroxymethyl rotation is almost two orders of magnitude higher than that of the molecular reorientation [41]. The results here demonstrate that this observation is consistent across the NPG derivatives.

The trend in activation energies show that the chemical modification of NPG has the largest impact on the hydroxymethyl rotation ($E_{a,2}$), which display the largest spread. This implies that the chemical perturbation of NPG is mostly strongly affecting the rotation related to the formation of the hydrogen bond network, which is facilitated by the hydroxymethyl functional groups. It also demonstrates the difference between lightly doped samples (NPG+PE, NPG+DND) compared to more significant chemical modification present in the solid solution (NPG+PG+PE), where the latter exhibits the largest relative change of $E_{a,2}$ and $E_{a,3}$ compared to pure NPG. The values for NPG are close to the values predicted for each mode from previous molecular dynamics studies [27]. The PC mode activation energies are shifted to lower absolute values than previously measured for NPG using QENS, due to the restricted $\Gamma_0$ values (though the relative magnitudes remain unchanged). Note that the small variation in measured methyl rotation activation energies ($E_{a,1}$) provides justification for implementing this method of fixing cross-sample attempt frequencies.

**Table 1.** Arrhenius parameters extracted from the global QENS+IFWS fits for the three resolved rotational modes in all four samples. The Arrhenius prefactors ($\Gamma_{0,i}$) are fixed at the pure-NPG best-fit values across all samples to make the relative activation energies ($E_{a,i}$) comparable. Relative changes in $E_a$ are calculated with respect to pure NPG.

| **Mode 1 (Methyl Rotation)** | | | |
|---|---|---|---|
| **Sample** | $\frac{1}{2}\Gamma_{0,1}$ (meV) | $E_{a,1}$ (meV) | Relative Change |
| **NPG** | $0.967 \pm 0.016$ | $98.4 \pm 0.3$ | - |
| **NPG+PE** | 0.967 (fixed) | $100.4 \pm 0.1$ | +2.0 % |
| **NPG+DND** | 0.967 (fixed) | $99.9 \pm 0.1$ | +1.5 % |
| **NPG+PG+PE** | 0.967 (fixed) | $99.5 \pm 0.1$ | +1.1 % |
| **Mode 2 (Hydroxymethyl Rotation)** | | | |
| **Sample** | $\frac{1}{2}\Gamma_{0,2}$ (meV) | $E_{a,2}$ (meV) | Relative Change |
| **NPG** | $394 \pm 146$ | $289.3 \pm 0.3$ | - |
| **NPG+PE** | 394 (fixed) | $283.4 \pm 0.3$ | -3.0 % |
| **NPG+DND** | 394 (fixed) | $278.8 \pm 0.4$ | -3.6 % |
| **NPG+PG+PE** | 394 (fixed) | $321.5 \pm 0.3$ | +11.1 % |
| **Mode 3 (Molecular Reorientation)** | | | |
| **Sample** | $\frac{1}{2}\Gamma_{0,3}$ (meV) | $E_{a,3}$ (meV) | Relative Change |
| **NPG** | $10.3 \pm 2.7$ | $239.1 \pm 0.3$ | - |
| **NPG+PE** | 10.3 (fixed) | $239.9 \pm 0.3$ | +0.3 % |
| **NPG+DND** | 10.3 (fixed) | $240.5 \pm 0.5$ | +0.6 % |
| **NPG+PG+PE** | 10.3 (fixed) | $255.1 \pm 0.3$ | +5.0 % |

The Arrhenius behaviour of the three modes presented in **Table 1** across the four samples is visualised in **Fig. 4**, where ln[Γ(meV)] is plotted against 1000/*T*, so the slope of each fitted line encodes the activation energy $E_a$ of the corresponding mode, and the intercept gives the fixed attempt-frequency pre-factors ln[$\Gamma_0$]. The vertical dashed line marks the OC-PC phase transition for each sample, and the dashed coloured lines are the Arrhenius extrapolations of the fitted Γ(*T*) for each mode with the shaded bands indicating the systematic uncertainty in $E_a$ propagated from the uncertainty in the fixed $\Gamma_0$ values.

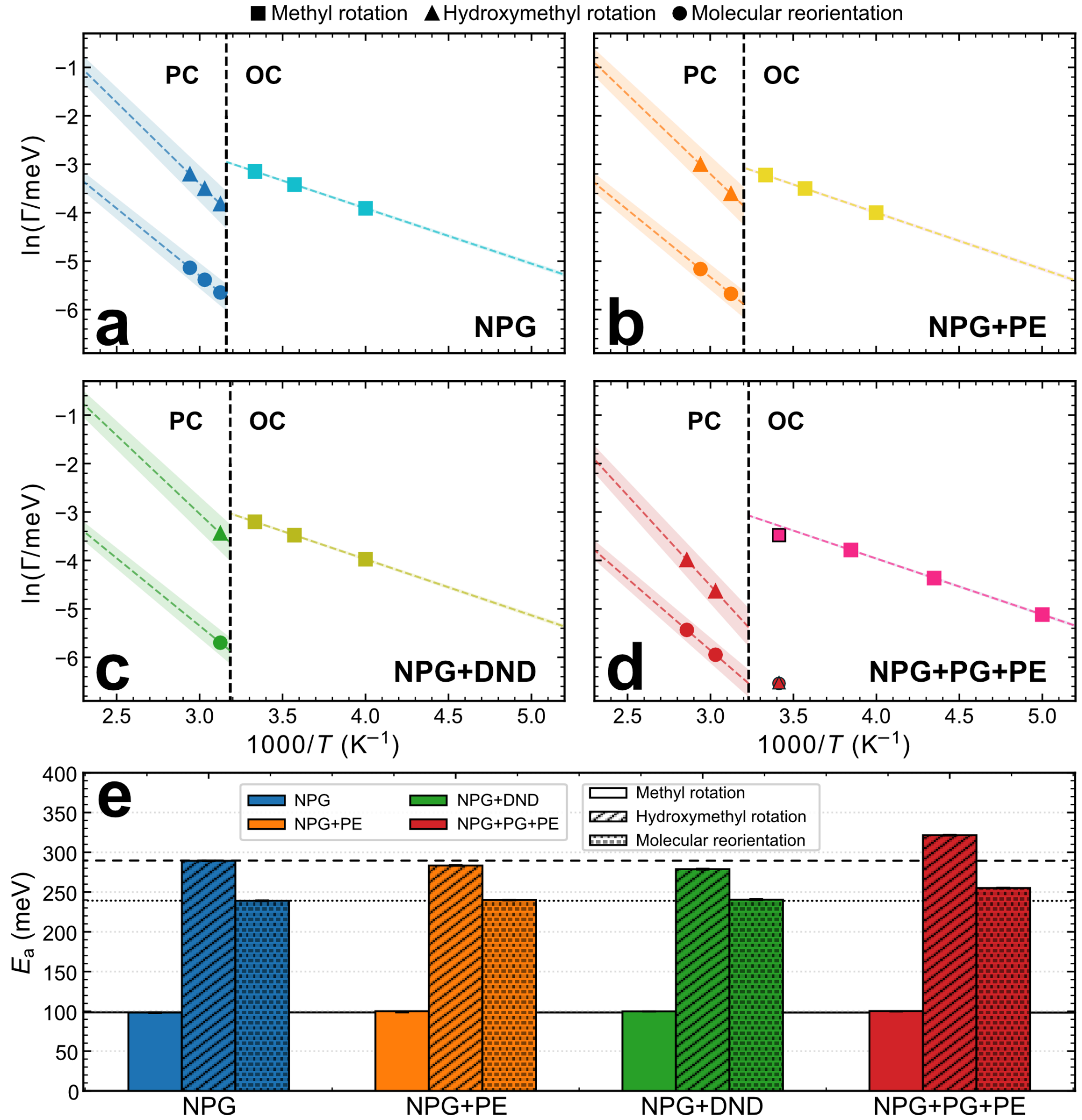

**Figure 4:** Arrhenius plots of the fitted Lorentzian linewidths Γ(*T*) for the three resolved rotational modes in (a) NPG, (b) NPG+PE, (c) NPG+DND, and (d) NPG+PG+PE, derived from global fits. The vertical black dashed lines give the OC to PC transition temperature. The dashed coloured lines are the Arrhenius extrapolations of the fitted Γ(*T*) with the shaded regions providing the systematic uncertainty in $E_a$, common to all samples. (e) Calculated activation energies, $E_a$, for each mode across the samples, where solid, dashed and dotted lines denote activation energies of methyl rotation, hydroxymethyl rotation and molecular reorientation, respectively, of NPG.

NPG, NPG+PE, and NPG+DND display broadly similar Arrhenius behaviour with some variation in the measured PC phase activation energies (**Fig. 4a-c**), while NPG+PG+PE shows substantially more pronounced variation in PC phase activation energies (**Fig.4d**). The markers with black outlines in **Fig. 4d** correspond to the additional QENS measurement at $T = 293$ K located within the hysteresis region of NPG+PG+PE, showing that OC methyl rotation persists alongside an additional QENS signal with a linewidth intermediate between the two PC modes. This is quite surprising, as it indicates that a large portion of the sample is still in the OC phase (with the corresponding methyl rotation frequency expected at this temperature), even though the PC modes are liberated enough to generate a detectable QENS signal. The QENS fit for this temperature is shown in **Supplementary Fig. S10d**, displaying both a high- and low-frequency mode unlike the previous QENS scan at 260 K (**Supplementary Fig. S10c**). This finding is consistent with a previous study of NPG+PG+PE, whereby the sample was shown to exhibit a very broad phase transition, with a disordered onset region before the bulk transition takes place [41]. **Figure 4e** summarises the calculated activation energies for each mode across the samples, with horizontal lines drawn at the pure-NPG values for ease of reference.

The relative PC phase activation-energy variations indicate two mechanistically distinct routes for perturbing the PC phase molecular dynamics. (i) NPG+PE and NPG+DND show a modest decrease in hydroxymethyl rotation activation energy with little change in the whole-molecule reorientation activation energy. This pattern is consistent with PE molecules and DND particles locally disrupting the inter-molecular hydrogen-bond network at substituted sites (lowering the hydroxymethyl rotational barrier), whilst having little effect on the tumbling of the molecules once in the PC phase. (ii) NPG+PG+PE exhibits a large variation across both PC modes: a significant increase in $E_{a,2}$ (+11.1%) and $E_{a,3}$ (+5.0%). This is perhaps not unexpected given that NPG+PG+PE has a much larger chemical variation and configurational disorder compared to the doped samples. NPG+PG+PE contains more hydroxymethyl groups than the other samples, which may account for the increase in both $E_{a,2}$ and $E_{a,3}$, signifying a more crowded hydroxy environment with more opportunities for hydrogen bonding. Together, the two signatures distinguish local hydrogen-bond network disruption (NPG+PE, NPG+DND) and configurational disorder (NPG+PG+PE) as distinguishable chemical-perturbation routes that QENS resolves directly.

## 3.2 Comparing Mobile Scatterers

The geometric character of the resolved rotational modes is encoded in the $Q$-dependence of the elastic incoherent structure factor (EISF). **Figure 5** shows the EISF for the two PC-phase modes, namely hydroxymethyl rotation (**Fig. 5a**) and molecular reorientation (**Fig. 5b**), each fitted with their respective geometric models. A discussion of the geometric forms of the EISFs and how the signals were separated is provided in **Supplementary Notes S1.3-S1.4**. In both cases the geometric radii are fixed, with $r_1$ = 0.94 Å, $r_2$ = 1.66 Å for the two-circle hydroxymethyl model and $r$ = 2.10 Å for the isotropic reorientation model. The quality of the fits in **Fig. 5** demonstrates the validity of the chosen geometric models. This leaves only the mobile fraction $f$ as a free fit parameter for comparison between samples.

The mobile fraction, $f$, included in the fitting of EISF data is attributed to the ratio of scatterers that are observed to be mobile on the timescale of the spectrometer at a particular temperature. A value less than unity implies that some portion of scatterers participating in a particular mode are immobile or otherwise sterically hindered due to their local environment [40]. For the PC-phase modes shown in **Fig. 5** it may also indicate a proportion of untransformed regions of the sample still in the OC phase. This would explain why $f$ increases with $T$ for hydroxymethyl rotation and molecular reorientation, as shown in **Supplementary Fig. S19**. The fact that $f$ is observed to be less than unity has potential implications for the accessible entropy change across the phase transition.

The temperatures chosen for **Fig. 5** are the first QENS scans after the OC-PC phase transition in the PC phase. This shows that NPG+PG+PE exhibits the most liberated hydroxymethyl rotation, whilst NPG+DND exhibits the most liberated molecular reorientation.  The fitted mobile fractions averaged over the measured temperatures are summarised in **Table 2**; their full temperature dependence is shown in **Supplementary Figs. S16-S19**.

Inspecting **Table 2**, the temperature-averaged OC methyl-rotation mobile fractions $\langle f_1 \rangle$ for NPG and NPG+DND cluster near 0.40-0.44, approaching the expected stoichiometric expectation that approximately one half of the hydrogen population, the six methyl hydrogens out of twelve per NPG molecule, participate in the methyl-rotation mode within the dynamic window. The other six hydrogens are involved in hydrogen-bonding in the OC phase and are thus static on the timescale of IN16B. The lower value for NPG+PG+PE (0.341) is expected as this reflects the reduced methyl-bearing molecular fraction in the ternary mixture, in which PG and PE replace a fraction of the NPG molecules and contribute fewer methyl groups. The low value for NPG+PE (0.327) is unexpected and may be due to residual signal near the elastic line not being included in the single-mode model, shown in **Supplementary Fig. S6**.

The temperature-averaged PC hydroxymethyl-rotation mobile fractions $\langle f_2 \rangle$ for NPG, NPG+PE, and NPG+DND lie in the range 0.76-0.80, with the ternary NPG+PG+PE showing a markedly higher value of 0.903. As with the reduced methyl fraction, this is consistent with the enhanced hydroxymethyl-group population in the ternary mixture: PG carries three hydroxymethyl groups per molecule and PE four, compared with two for NPG. The combination of increased hydroxymethyl population and the elevated $E_{a,2}$ for NPG+PG+PE (**Table 1**) is consistent with a more populated but crowded hydroxymethyl rotational landscape. All four samples display similar sensitivity of $f_2$ with increasing temperature, as shown in **Supplementary Fig. S19**. The temperature-averaged PC molecular-reorientation mobile fractions $\langle f_3 \rangle$ cluster near 0.83-0.88 across all four samples, however there is a strong sensitivity of $f_3$ with temperature for this mode, as shown in **Supplementary Figs. S18-S19**. This suggests that the mobility of the molecular tumbling mode is not fully liberated immediately after the OC-PC phase transition in all samples. Furthermore, NPG+PE specifically displays the largest increase in both $f_2$ and $f_3$ with increasing temperature, indicating it may exhibit strong reversible BC effects, since the mobility of these modes are linked to the accessible entropy change through the BC phase transition [27,40].

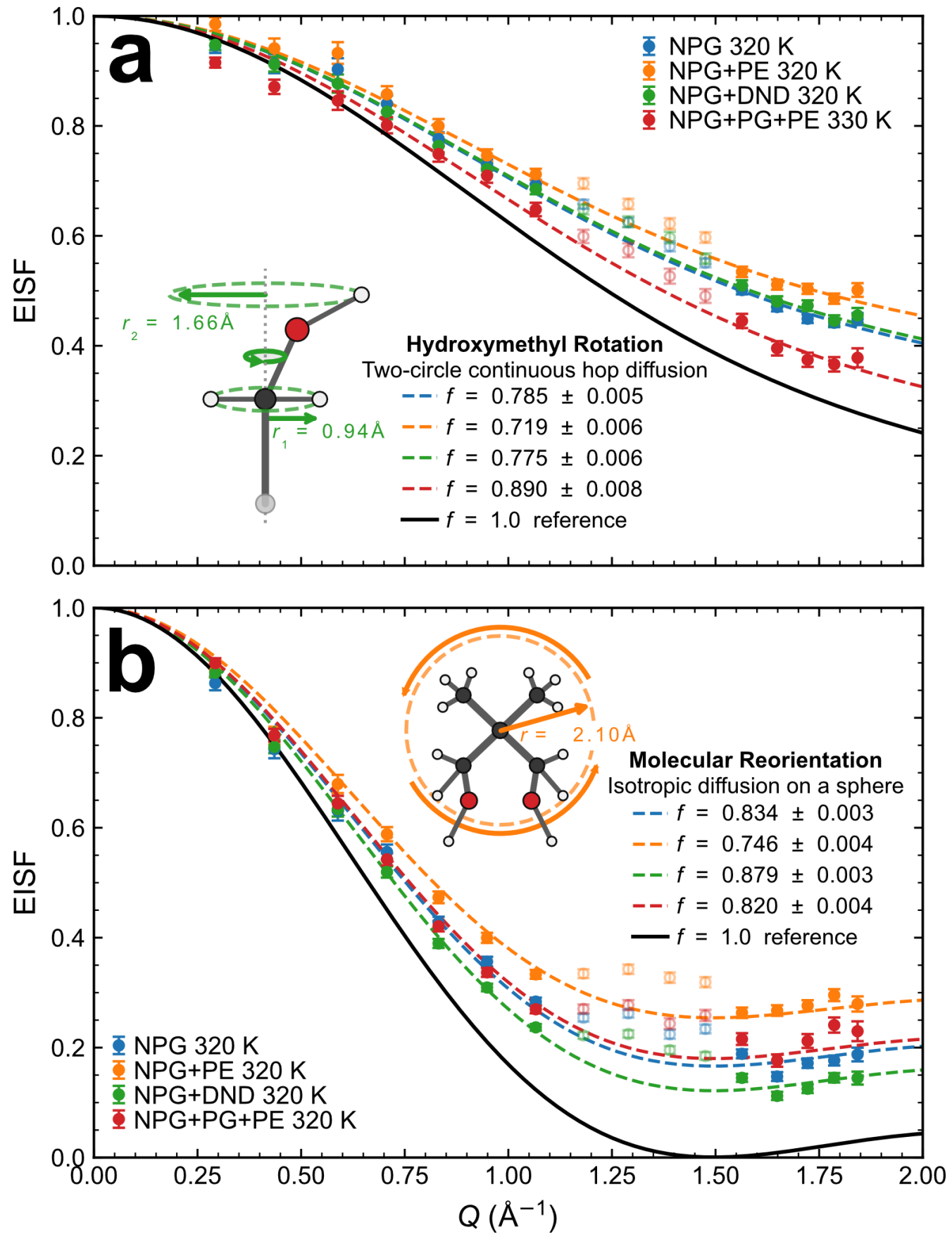

**Figure 5:** Comparison of the elastic incoherent structure factor (EISF) for the two PC-phase rotational modes across all samples, fitted with their respective geometric models. (a) EISF for hydroxymethyl rotation, fitted with the two-circle continuous diffusion model. (b) EISF for molecular reorientation, fitted with isotropic continuous diffusion on a sphere. In both panels the only free fit parameter is the mobile fraction *f* and the solid black curve shows the respective $f = 1.0$ reference (fully mobile limit). Dashed lines are the best-fit curves for each sample. Open symbols are excluded points due to Bragg peaks in this *Q*-range.

These findings indicate that the population of scatterers involved in hydroxymethyl rotation and whole-molecule reorientation are sensitive to the chemical perturbations explored here. Combined with the activation-energy variations of **Table 1**, we suggest that ($\langle f \rangle$, $E_a$) values provide a per mode fingerprint that distinguishes chemical-perturbation routes more sensitively than either quantity alone.

**Table 2.** Temperature-averaged EISF mobile fractions ⟨*f*⟩ obtained from the geometric model fits across all samples. $\langle f_1 \rangle$ refers to OC methyl rotation; $\langle f_2 \rangle$ to PC hydroxymethyl rotation; and $\langle f_3 \rangle$ to PC molecular reorientation. Uncertainties are standard deviations across the included temperatures (see **Supplementary Figs. S16-S18**).

| Sample | $\langle f_1 \rangle$ (OC, methyl rotation) | $\langle f_2 \rangle$ (PC, hydroxymethyl rotation) | $\langle f_3 \rangle$ (PC, molecular reorientation) |
|---|---|---|---|
| **NPG** | 0.402 ± 0.025 | 0.803 ± 0.017 | 0.877 ± 0.040 |
| **NPG + PE** | 0.327 ± 0.032 | 0.760 ± 0.058 | 0.830 ± 0.118 |
| **NPG + DND** | 0.446 ± 0.005 | 0.775 (single *T*) | 0.879 (single *T*) |
| **NPG + PG + PE** | 0.341 ± 0.003 | 0.903 ± 0.018 | 0.847 ± 0.037 |

Furthermore, the EISF geometric radii and the PC-phase $\Gamma_0$ values are deliberately fixed across all samples; this trades absolute accuracy for cross-sample comparability, which is the central design choice of this methodology rather than an attempt to extract definitive single-sample parameters. None of these constraints affects the central finding: that samples can be distinguished by characteristic mobility (*f*) and activation energy ($E_a$) parameters, but it does mean that the absolute values reported here should be compared with appropriate caution to those obtained by other methods or under different fitting assumptions.

# 4 Conclusions

We have developed a comparative QENS+IFWS global-fitting methodology for the molecular dynamics of BC molecular crystals and demonstrated it on NPG and three chemically distinct derivatives. By constraining the attempt frequencies to their pure-NPG values across samples, we obtain statistically precise relative activation energies and EISF-derived mobile fractions across three distinct rotational modes. The three derivatives studied demonstrate two mechanistically distinct routes for perturbing the molecular dynamics that underpin the order-disorder transition. NPG+PE and NPG+DND appear to locally disrupt the hydrogen-bond network resulting in a ~3% reduction in the hydroxymethyl rotation activation energy. NPG+PG+PE introduces significant configurational disorder that increases the orientational potential, likely due to the increased hydroxymethyl group population. The framework presented here may be generalised to other plastic crystal or molecular systems in which the constituent molecule(s) support distinguishable rotational modes accessible to high-resolution backscattering neutron spectroscopy.

## Data accessibility

All data underpinning this work are held on the ILL database (https://doi.ill.fr/10.5291/ILL-DATA.7-02-221, https://doi.ill.fr/10.5291/ILL-DATA.7-02-227, and https://doi.ill.fr/10.5291/ILL-DATA.EASY-1525).

## Declaration of AI use

The AI model Claude Opus 4.7 was used to assist the writing of Python code to plot the figures in this work, and to assist in improving the language and readability of the prose. AI was not used to analyse or interpret any of the data presented in this work.

## Author's contributions

**F.R.B**.: Conceptualization, Formal analysis, Funding acquisition, Investigation, Methodology, Visualization, Writing – original draft, Writing – review & editing. **D.B.**: Funding acquisition, Investigation, Supervision, Writing – review & editing. **M.A.**: Formal analysis, Investigation, Methodology, Supervision, Writing – review & editing. **D.A.M.**: Funding acquisition, Investigation, Supervision, Writing – review & editing.

## Conflict of interest declaration

The authors declare no conflict of interest.

## Funding

This work was supported by a Leverhulme Early Career Research Fellowship (ECF-2025-577) and an EPSRC grant (EP/V042262/1).

## Acknowledgements

The authors acknowledge beam time awarded by the ILL neutron source (proposals 7-02-221, 7-02-227 and EASY-1525).

# Supplementary Information:

# Insight into the molecular dynamics of barocaloric molecular crystals using quasi-elastic neutron scattering

Frederic Rendell-Bhatti[1,*], David Boldrin[1], Markus Appel[2] and Donald A. MacLaren[1]

[1]SUPA, School of Physics and Astronomy, University of Glasgow, Glasgow, UK

[2]Institut Laue-Langevin (ILL), 71 Av. Des Martyrs, 38000 Grenoble, France

*Email Correspondence: fred.rendell@glasgow.ac.uk

## Supplementary Note S1: QENS+IFWS global fitting analysis

### S1.1 Arrhenius parameterisation and global fitting

The dynamic structure factor measured in quasi-elastic neutron scattering (QENS) is a function of momentum (*Q*) and energy (*E*) transfer and is modelled as:

$$S(Q,E) = [A_0(Q)\delta(E) + \Sigma_i A_i(Q) L_i(E)] \otimes R(Q,E) + B(Q), \quad \text{(S1)}$$

where $A_0(Q)$ is the amplitude of the elastic peak, representing scattering from atoms immobile within the instrumental dynamic window. The sum $\Sigma_i$ over *n* Lorentzian components captures the distinct quasi-elastic modes resolved at a given temperature, each with amplitude $A_i(Q)$ and full-width at half maximum (FWHM) linewidth $\Gamma_i$. $R(Q,E)$ is the experimentally determined instrumental resolution function and $B(Q)$ is a *Q*-dependent, energy-independent flat background term. Each Lorentzian takes the form:

$$L_i = \frac{1}{\pi} \cdot \frac{\frac{1}{2}\Gamma_i(Q)}{E^2 + (\frac{1}{2}\Gamma_i(Q))^2}. \quad \text{(S2)}$$

The inelastic fixed window scan (IFWS) measures integrated scattering intensity at a fixed non-zero energy transfer, for example $E_{\text{off}} = \pm 3$ μeV, as a function of temperature. For an individual Lorentzian mode *i*, the IFWS signal at a particular *Q* is given by:

$$S_{\text{IFWS},i}(E_{\text{off}}, T) = \frac{A_i}{\pi} \cdot \frac{\frac{1}{2}\Gamma_i(T)}{{E_{\text{off}}}^2 + (\frac{1}{2}\Gamma_i(T))^2}. \quad \text{(S3)}$$

The temperature, $T'$, at which the IFWS intensity peak occurs satisfies $\Gamma_i(T') \approx 2E_{\text{off}}$, providing a characteristic temperature signature for each mode that is set by its Arrhenius parameters. When multiple modes are measured simultaneously in the PC phase, the total IFWS signal is:

$$S_{\text{IFWS}}(E_{\text{off}}, T) = \Sigma_i S_{\text{IFWS},i}(E_{\text{off}}, T) + B(T), \quad \text{(S4)}$$

with each mode contributing its own peak at a characteristic $T'$, as shown in Figure 3d of the manuscript. In addition to the IFWS intensity coming from the modes detected in QENS, a flat background term $B(T)$ is also included in the IFWS fits.

The temperature dependence of each Lorentzian FWHM linewidth, $\Gamma_i(T)$, is assumed to follow an Arrhenius relationship:

$$\Gamma_i(T) = \Gamma_{0,i} \cdot e^{\frac{-E_{a,i}}{k_B T}}, \quad \text{(S5)}$$

where $\Gamma_{0,i}$ is the prefactor of mode *i*, $E_{a,i}$ is the activation energy of mode *i*, and $k_B$ is the Boltzmann constant. Taking logarithms gives the linearised form:

$$\ln[\Gamma_i(T)] = \ln(\Gamma_{0,i}) - \frac{E_{a,i}}{k_B T}, \quad \text{(S6)}$$

which is linear in $\frac{1}{T}$ with slope $(-\frac{E_{a,i}}{k_B})$ and intercept $\ln(\Gamma_{0,i})$.

The global fit function minimises a single $\chi^2$ over all QENS spectra (at the fitted temperatures) and all IFWS data points (continuously sampled in $T$), under the constraint that the QENS Lorentzians and the IFWS contributions for each mode share a single set of $Q$-independent Arrhenius parameters ($\Gamma_{0,i}$, $E_{a,i}$) as given in Eq. S5.

Practically, this means that for each mode *i*, the $\Gamma_i(T)$ from equation S5 is tied between both the QENS model (Eq. S1, S2) at each measured temperature and the IFWS model (Eq. S3) at the continuous IFWS temperatures. Furthermore, for temperatures where two modes are present, an additional tie is used between the QENS and IFWS datasets to ensure identical ratios of the amplitudes of Lorentzian contributions $\frac{A_1(Q)}{A_2(Q)}$ across all spectra with the same $Q$.

## S1.2 $\Gamma_0$–$E_a$ correlation

The Arrhenius parameters $\Gamma_0$ and $E_a$ are strongly correlated when fitted over a limited temperature range, and especially when multiple modes are present in the high-temperature plastic crystal (PC) phase. From Eq. S6, at any single temperature $T$, a variation in $\ln(\Gamma_0)$ can be compensated by a corresponding variation in $E_a$, that leaves $\Gamma(T)$ unchanged.

To account for these highly correlated parameters, especially in the PC phase with a limited ~45 K IFWS temperature range and two modes present, pure NPG was initially globally fit allowing $\Gamma_{0,i}$ and $E_{a,i}$ to refine to their best-fit values for each mode, *i*. This resulted in methyl rotation (Mode 1, *i* = 1): $\frac{\Gamma_{0,1}}{2} = 0.967 \pm 0.016\ \text{meV}$, hydroxymethyl rotation (Mode 2, *i* = 2): $\frac{\Gamma_{0,2}}{2} = 394 \pm 146\ \text{meV}$ and molecular reorientation (Mode 3, *i* = 3): $\frac{\Gamma_{0,3}}{2} = 10.3 \pm 2.72\ \text{meV}$. These best-fit values were then fixed in all future global fits for all samples, to obtain relative $E_a$ values with a high statistical confidence. Crucially, the systematic uncertainties are common to all samples when $\Gamma_{0,i}$ is held fixed at the same value across samples. Sliding the chosen $\Gamma_{0,i}$ value within its uncertainty range shifts all $E_{a,i}$ values together by the same amount, preserving their relative differences.

Thus, the following strategy was adopted for the QENS+IFWS global fitting as presented in the manuscript:

- All three modes inherit $\Gamma_{0,1}$, $\Gamma_{0,2}$ and $\Gamma_{0,3}$ as fixed values from the pure-NPG best-fit values across all samples. The cross-sample variation is then captured entirely in the $E_a$ values, which are determined with high statistical precision from the data. The shaded regions in Figure 4a-d of the main text represent the systematic uncertainties in $E_a$ (±0.5 meV for Mode 1, ±10.2 meV for Mode 2 and ±7.3 meV for Mode 3) arising from the original $\Gamma_{0,i}$ uncertainty.

## S1.3 Geometric EISF models

Each rotational mode is associated with a geometric model that determines the $Q$-dependence of its elastic incoherent structure factor (EISF):

$$\mathrm{EISF}_i(Q) = (1 - f_i) + f_i \cdot \mathrm{EISF}_i^{\mathrm{geom}}(Q), \tag{S7}$$

where $f_i$ is the fraction of mobile hydrogen scatterers for mode $i$ within the IN16B dynamic window and $\mathrm{EISF}_i^{\mathrm{geom}}(Q)$ is the geometric structure factor for the assumed motion. The three geometric forms used in the manuscript are as follows:

Mode 1 describes three-fold jump diffusion for methyl rotation. The methyl group is modelled as a discrete 3-site jump on a circle of radius $r$ = 0.92 Å [1], the perpendicular distance from the C–$CH_3$ rotation axis to each hydrogen as shown in Fig. 1e of the manuscript:

$$\mathrm{EISF}_1^{\mathrm{three-hop}}(Q) = \frac{1}{3}\left(1 + \frac{2\sin(\sqrt{3}Qr)}{\sqrt{3}Qr}\right). \tag{S8}$$

Mode 2 describes two-circle continuous diffusion for hydroxymethyl rotation. Unlike the methyl group, the hydroxymethyl group does not have 3-fold rotational symmetry around the rotation axis, making the three-fold jump model inappropriate. Instead, the hydroxymethyl group is modelled as a stoichiometric average of two continuous diffusion on a circle EISFs, with the $CH_2$ hydrogens on an inner circle of radius $r_1$ = 0.94 Å and the OH hydrogen with a mean outer circle radius of $r_2$ = 1.66 Å [2], as shown in Fig. 1f of the manuscript. The single-circle EISF for a six-site equally-spaced jump diffusion on a circle of radius $r$, which approximates continuous rotation [1], is:

$$\mathrm{EISF}^{\mathrm{circle}}(Q,r) = \frac{1}{6}\left(1 + \frac{2\sin(Qr)}{Qr} + \frac{2\sin(\sqrt{3}Qr)}{\sqrt{3}Qr} + \frac{\sin(2Qr)}{2Qr}\right). \tag{S9}$$

The combined hydroxymethyl EISF is then:

$$\mathrm{EISF}_2^{\mathrm{two-circle}}(Q) = \frac{2}{3}\mathrm{EISF}^{\mathrm{circle}}(Q,r_1) + \frac{1}{3}\mathrm{EISF}^{\mathrm{circle}}(Q,r_2), \tag{S10}$$

where the weights $\frac{2}{3}$ and $\frac{1}{3}$ reflect the stoichiometric ratio of $CH_2$ hydrogens to OH hydrogens.

Mode 3 describes isotropic diffusion on a sphere for molecular reorientation. Whole-molecule reorientation is modelled as continuous isotropic diffusion on a sphere of effective radius $r$ [1]:

$$\mathrm{EISF}_3^{\mathrm{diff-sphere}}(Q) = \left(\frac{\sin(Qr)}{Qr}\right)^2, \tag{S11}$$

where $r$ was determined by a pure NPG temperature-averaged, free-fit value and subsequently fixed at $r$ = 2.10 Å for all materials. This value is within about 5% of the value expected from the weighted average hydrogen distance from the centre of mass of the molecule, representing full isotropic molecular tumbling [2].

## S1.4 Mode-resolved EISF extraction from multi-Lorentzian fits

At each $Q$ value, the multi-Lorentzian fit in the PC phase returns amplitudes for the elastic peak and each Lorentzian. The per-mode experimental EISFs are then defined as ratios of these fitted amplitudes.

In the OC phase there is one Lorentzian, methyl rotation (Mode 1) with amplitude $A_{1,CH3}(Q)$ and EISF:

$$\mathrm{EISF}_1^{\mathrm{exp}}(Q) = \frac{A_0(Q)}{A_0(Q)+A_{1,CH3}(Q)}. \tag{S12}$$

In the PC phase there are two Lorentzians, hydroxymethyl rotation (Mode 2) with amplitude $A_{1,\mathrm{CH2OH}}(Q)$ and EISF:

$$\mathrm{EISF}_2^{\mathrm{exp}}(Q) = \frac{A_0(Q) + A_{1,mol}(Q)}{A_0(Q)+A_{1,CH2OH}(Q)+A_{1,mol}(Q)}, \tag{S13}$$

and molecular reorientation (Mode 3) with amplitude $A_{1,\mathrm{mol}}(Q)$ and EISF:

$$\mathrm{EISF}_3^{\mathrm{exp}}(Q) = \frac{A_0(Q)}{A_0(Q) + A_{1,mol}(Q)}. \tag{S14}$$

The separation of EISFs shown in Eqs. S13 and S14 are approximate, and valid when the timescales of the two modes are well-separated, as they are for hydroxymethyl rotation ($\tau_2 \approx 60$ ps) and molecular reorientation ($\tau_3 \approx 370$ ps), calculated for NPG at 320 K. This approximation ignores any contribution from the fast methyl rotation in the PC phase, which has a FWHM linewidth greater than the dynamic range of IN16B ($\tau_1 \approx 3$ ps) [2,3].

## S1.5 Flat background term

The included flat background term for both the QENS and IFWS fits can be justified as having two distinct contributions:

1. Fast methyl rotation in the PC phase. The methyl rotation operates throughout both phases of NPG, but in the PC phase its activation energy is reduced and its linewidth at the typical measurement temperatures (320–340 K) reaches approximately 0.5 meV [2], which is well outside the IN16B dynamic range of ±28 μeV. The wings of this very broad Lorentzian appear essentially flat across the measured energy window, contributing an approximately energy-independent intensity that depends weakly on *Q* through the methyl-rotation EISF.
2. Phonon contributions. At the measurement temperatures, scattering from low-energy phonon modes adds to the inelastic intensity. The energy distribution of these contributions is much broader than the IN16B window and appears effectively flat over the measured energy range.

A measurable *Q*-dependence of *B*(*Q*) was observed from the fit at each *Q* bin, and thus it is reasonable to expect that the fast PC phase methyl rotation is the strongest contributor to the flat background signal.

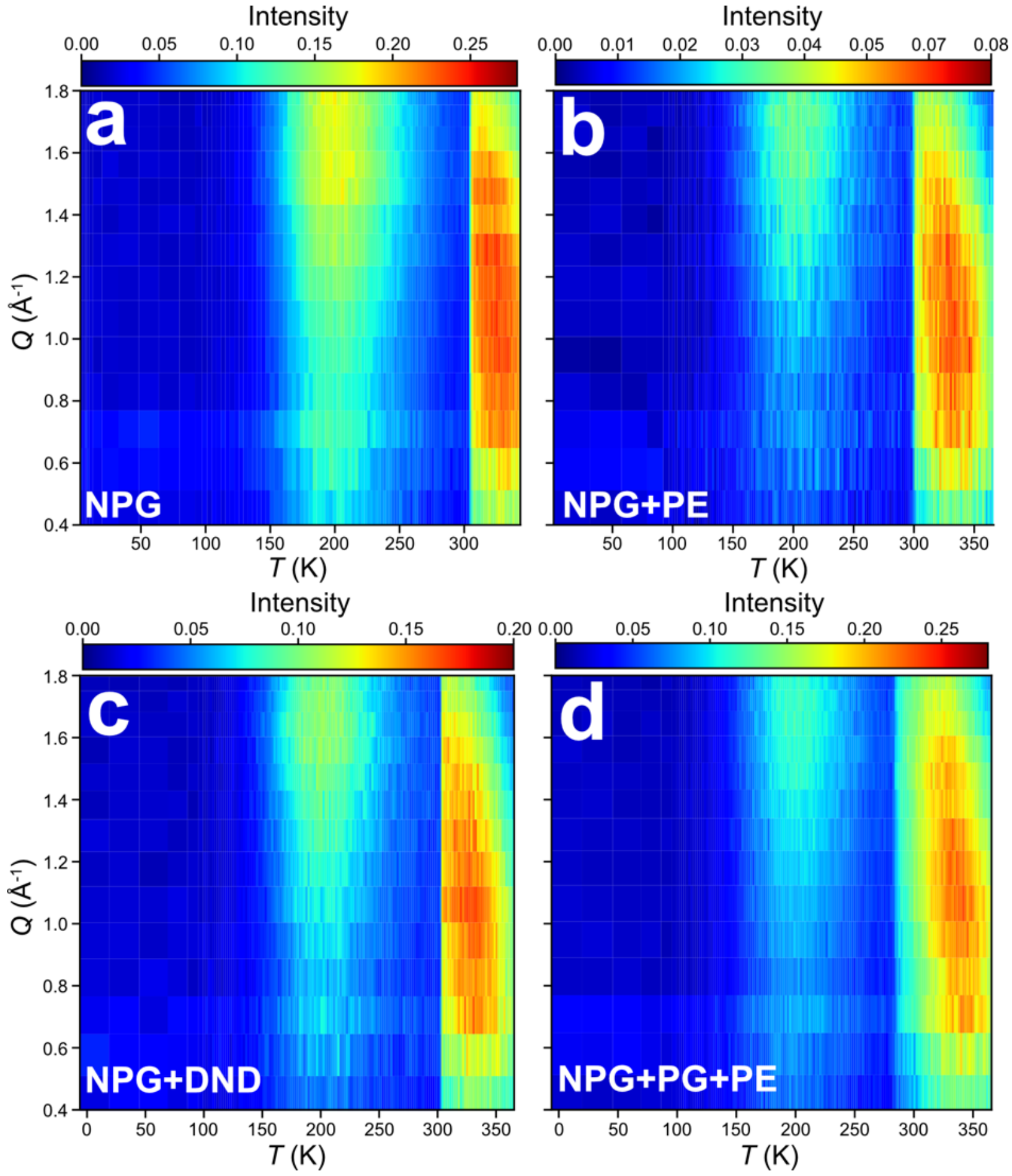

**Figure S1**: Raw IFWS ($E = \pm 3$ μeV, 0.4 Å$^{-1}$ < $Q$ < 1.8 Å$^{-1}$) on cooling for (a) NPG, (b) NPG+PE, (c) NPG+DND, (d) NPG+PG+PE. All samples show broadly the same features; a broad peak centred around 200 K corresponding to the OC phase methyl rotation and a sharp increase in intensity around 300 K corresponding to the liberation of the PC phase hydroxymethyl rotation and molecular reorientation modes.

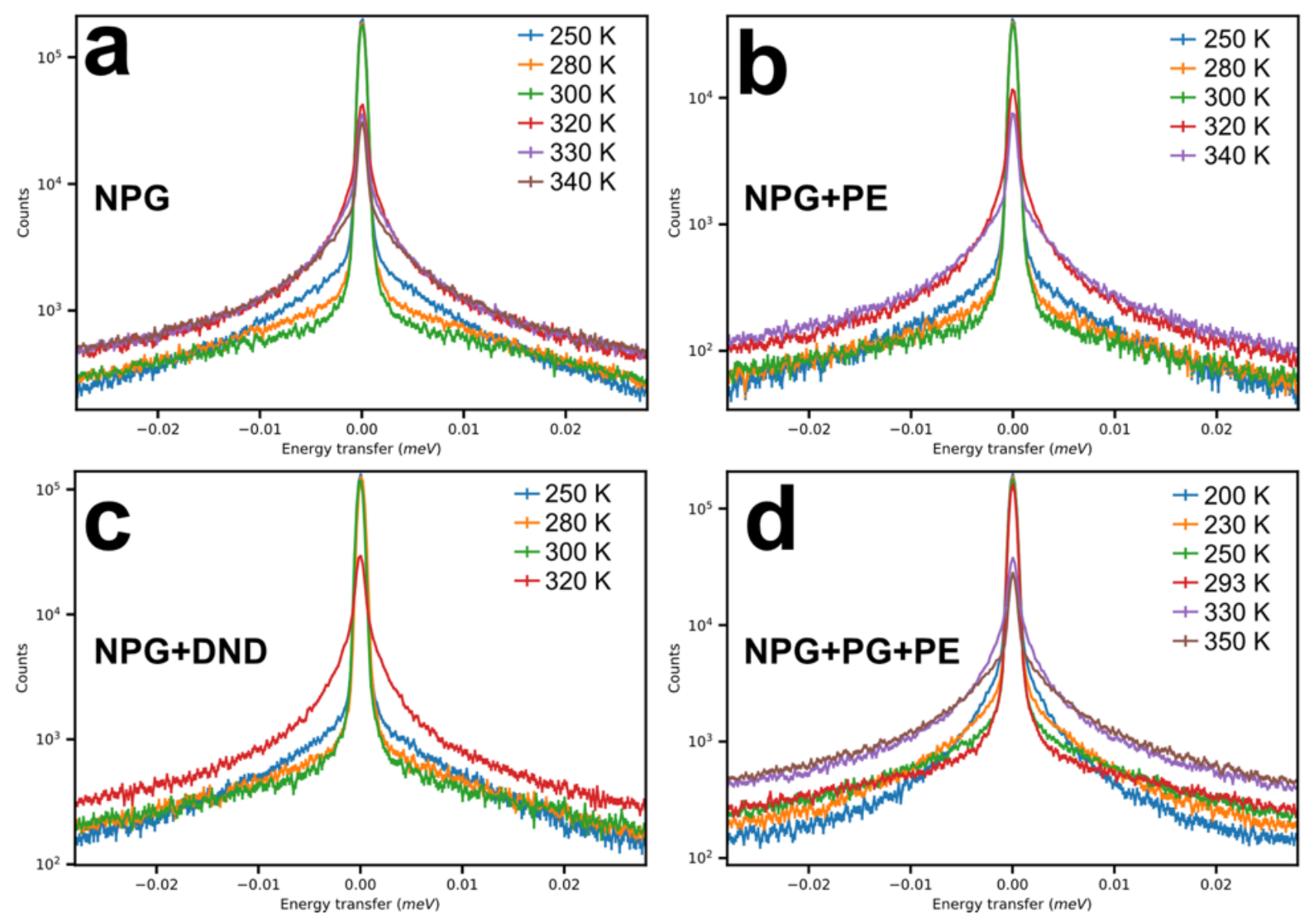

**Figure S2**: Raw QENS (0.19 Å$^{-1}$ < $Q$ < 1.89 Å$^{-1}$) for (a) NPG, (b) NPG+PE, (c) NPG+DND, (d) NPG+PG+PE. Ordered crystal (OC) phase displays a broadening of the QENS signal associated with methyl rotation on heating towards 300 K. Above 300 K all samples display a significant increase(decrease) in quasi-elastic(elastic) signal, consistent with the liberation of hydroxymethyl rotation and molecular reorientation modes. (d) An additional temperature ($T$ = 293 K) in the hysteresis region was obtained for NPG+PG+PE, and this QENS scan was not included in the global fit due to the presence of a mixture of OC and PC modes. This QENS data was instead fitted separately and compared to the globally fitted datasets, as presented in Fig. 4 of the main manuscript.

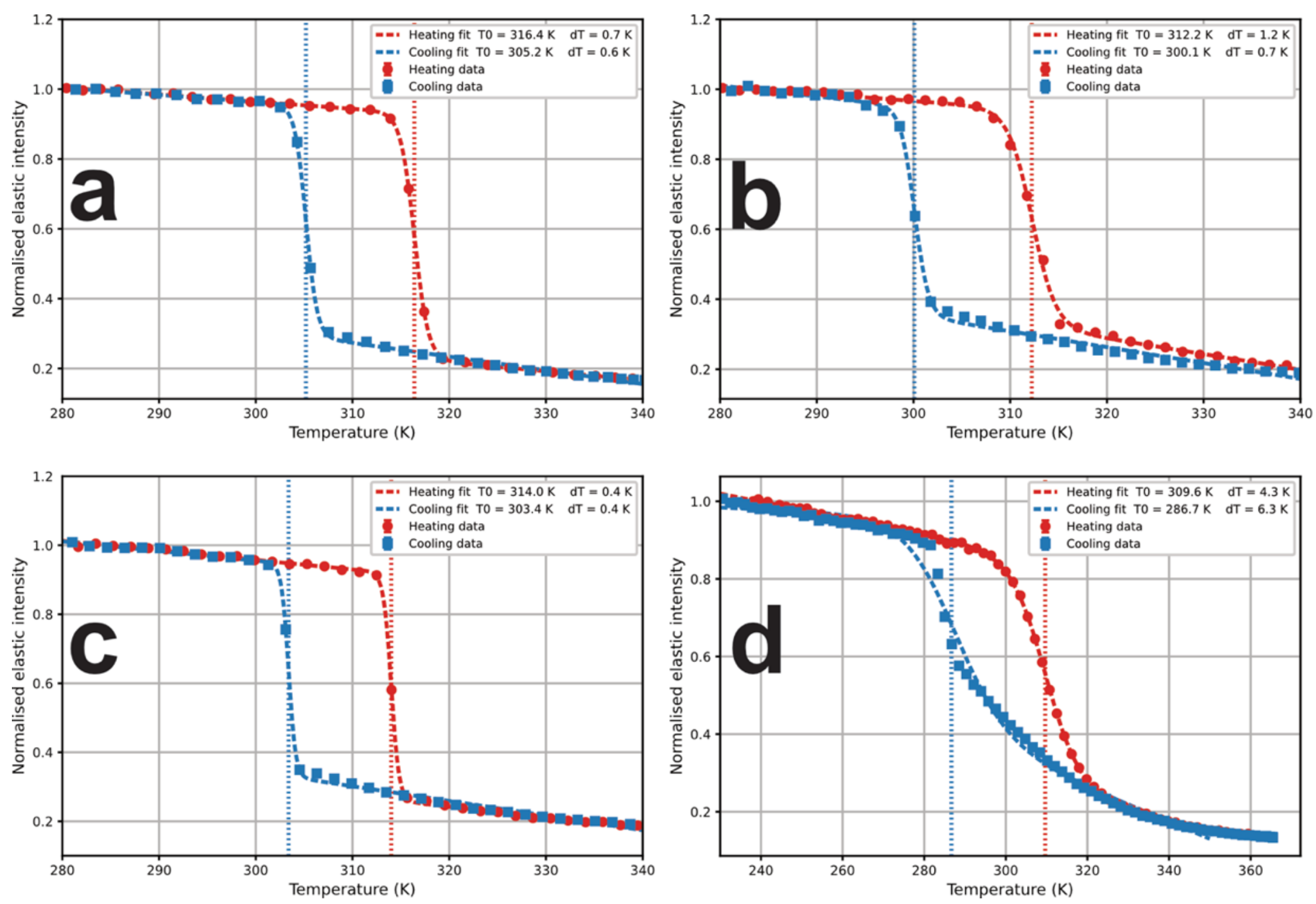

**Figure S3**: Elastic fixed-window scan (EFWS) on heating and cooling through the OC-PC phase transition, showing the hysteresis in molecular dynamics for (a) NPG, (b) NPG+PE, (c) NPG+DND, (d) NPG+PG+PE. The dashed line is a phenomenological Boltzmann sigmoid fit, using a linear baseline for either side of the phase transition. (a) NPG and (c) NPG+DND show sharp and square-like hysteresis behaviour, well modelled by the sigmoid fit, whereas (b) NPG+PE and (d) NPG+PG+PE exhibit more asymmetric hysteresis behaviour, which is particularly accentuated in NPG+PG+PE.

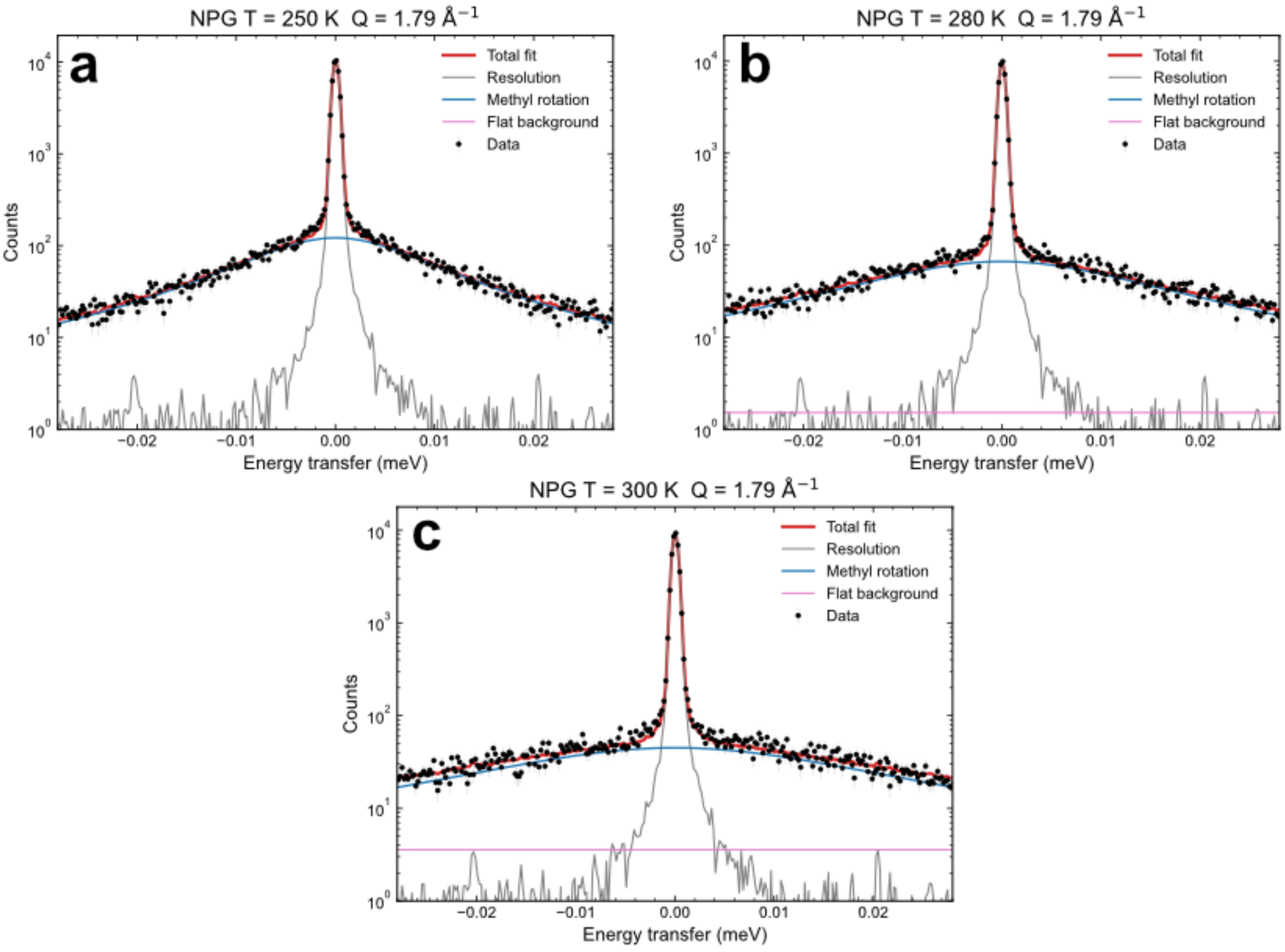

**Figure S4**: Representative QENS fits ($Q$ = 1.79 Å$^{-1}$) for NPG in the OC phase at (a) 250 K, (b) 280 K and (c) 300 K. Only a single Lorentzian is included in the model at these temperatures, accounting for the methyl rotation. The flat background term accounts for multiple-scattering and other inelastic scattering contributions.

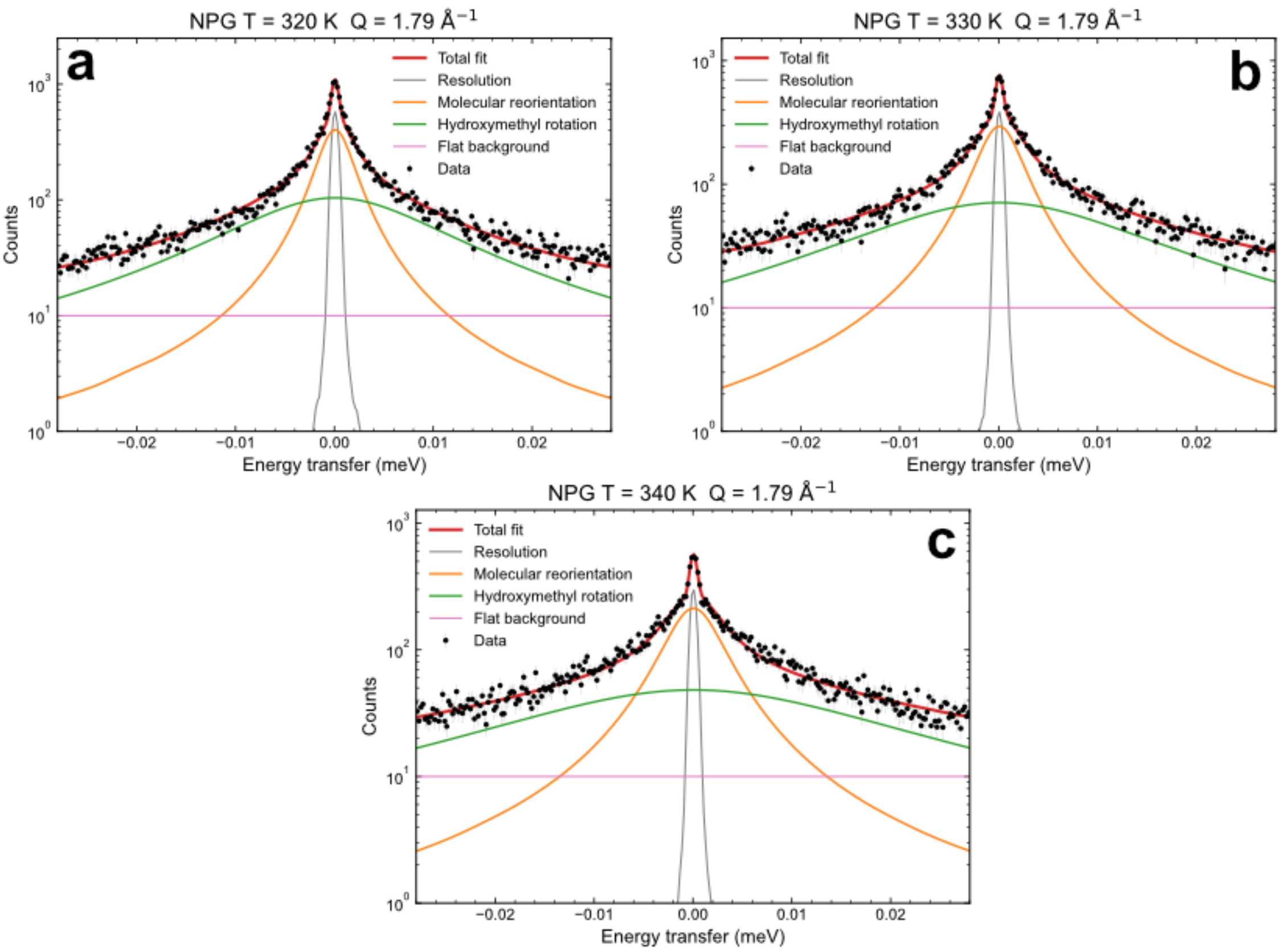

**Figure S5**: Representative QENS fits ($Q$ = 1.79 Å$^{-1}$) for NPG in the PC phase at (a) 320 K, (b) 330 K and (c) 340 K. Two Lorentzians are included in the model at these temperatures, accounting for the hydroxymethyl rotation and molecular reorientation. Methyl rotation is present in the PC phase but is not observed as part of the QENS signal as a Lorentzian component, since this mode is outside the dynamic range of the instrument and is therefore included in the flat background.

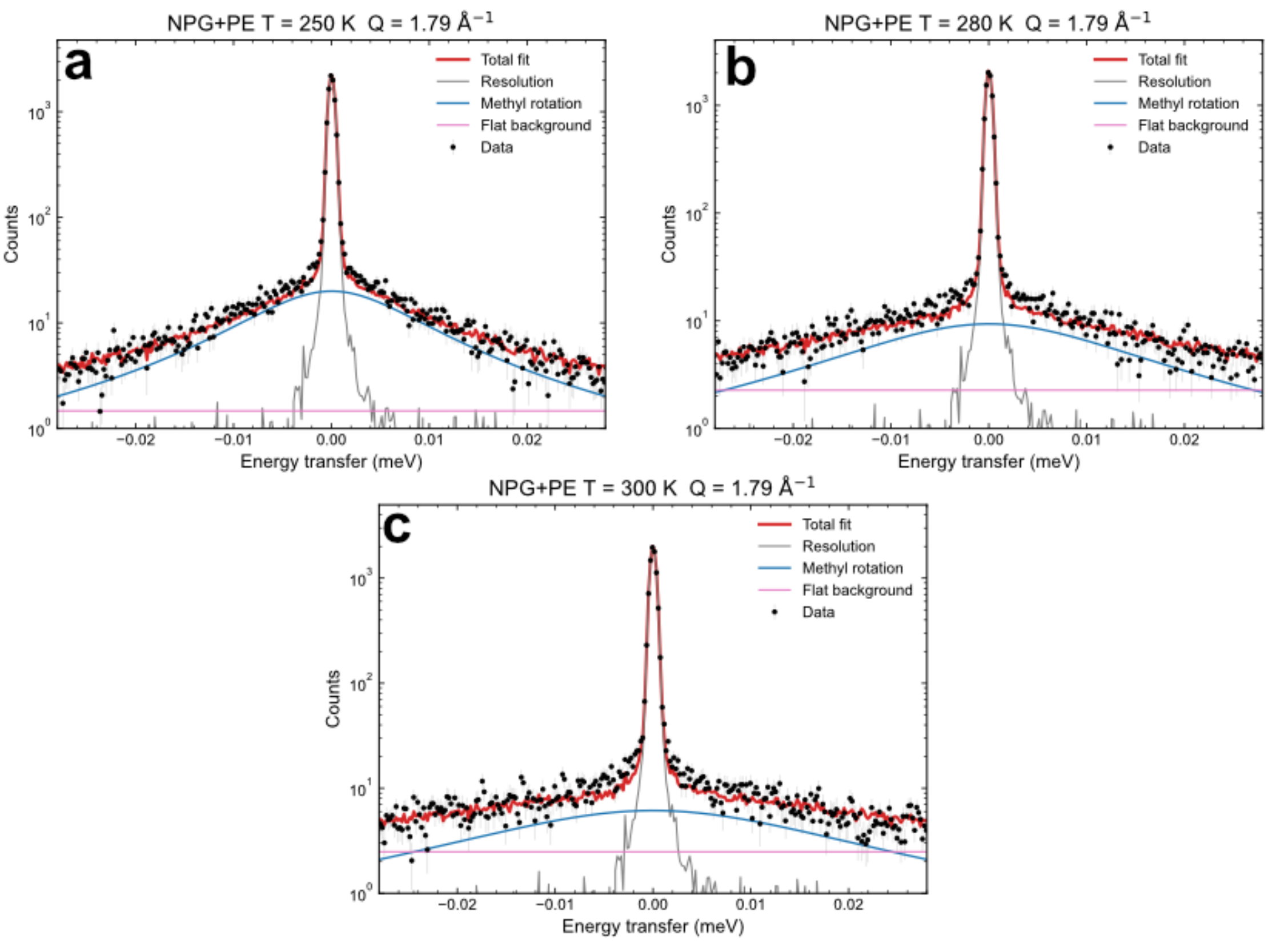

**Figure S6**: Representative QENS fits ($Q$ = 1.79 Å$^{-1}$) for NPG+PE in the OC phase at (a) 250 K, (b) 280 K and (c) 300 K. Only a single Lorentzian is included in the model at these temperatures, accounting for the methyl rotation. The flat background term accounts for multiple-scattering and other inelastic scattering contributions. There appears to be some residual signal at very low energy transfers, especially at 280 K and 300 K, suggesting the early onset of the PC rotational modes.

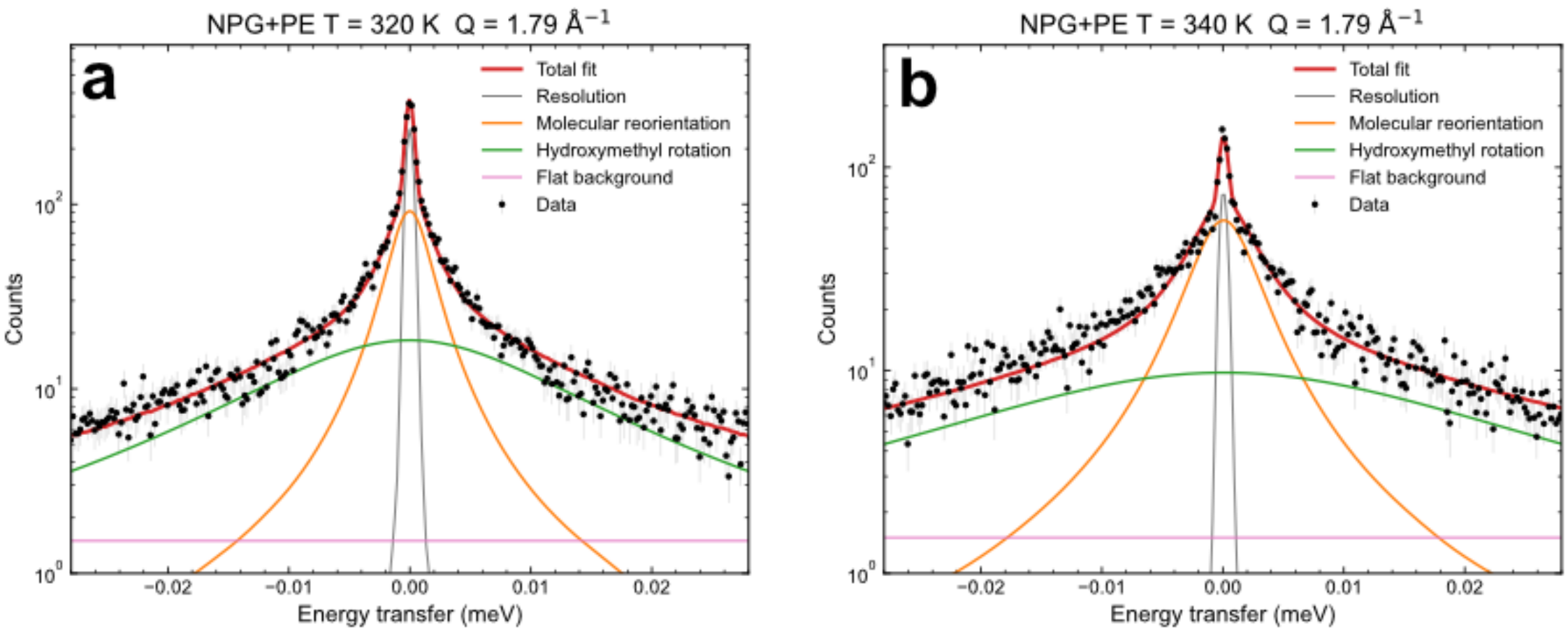

**Figure S7**: Representative QENS fits ($Q$ = 1.79 Å$^{-1}$) for NPG+PE in the PC phase at (a) 320 K, (b) 340 K. Two Lorentzians are included in the model at these temperatures, accounting for the hydroxymethyl rotation and molecular reorientation. Methyl rotation is present in the PC phase but is not observed as part of the QENS signal as a Lorentzian component, since this mode is outside the dynamic range of the instrument and is therefore included in the flat background.

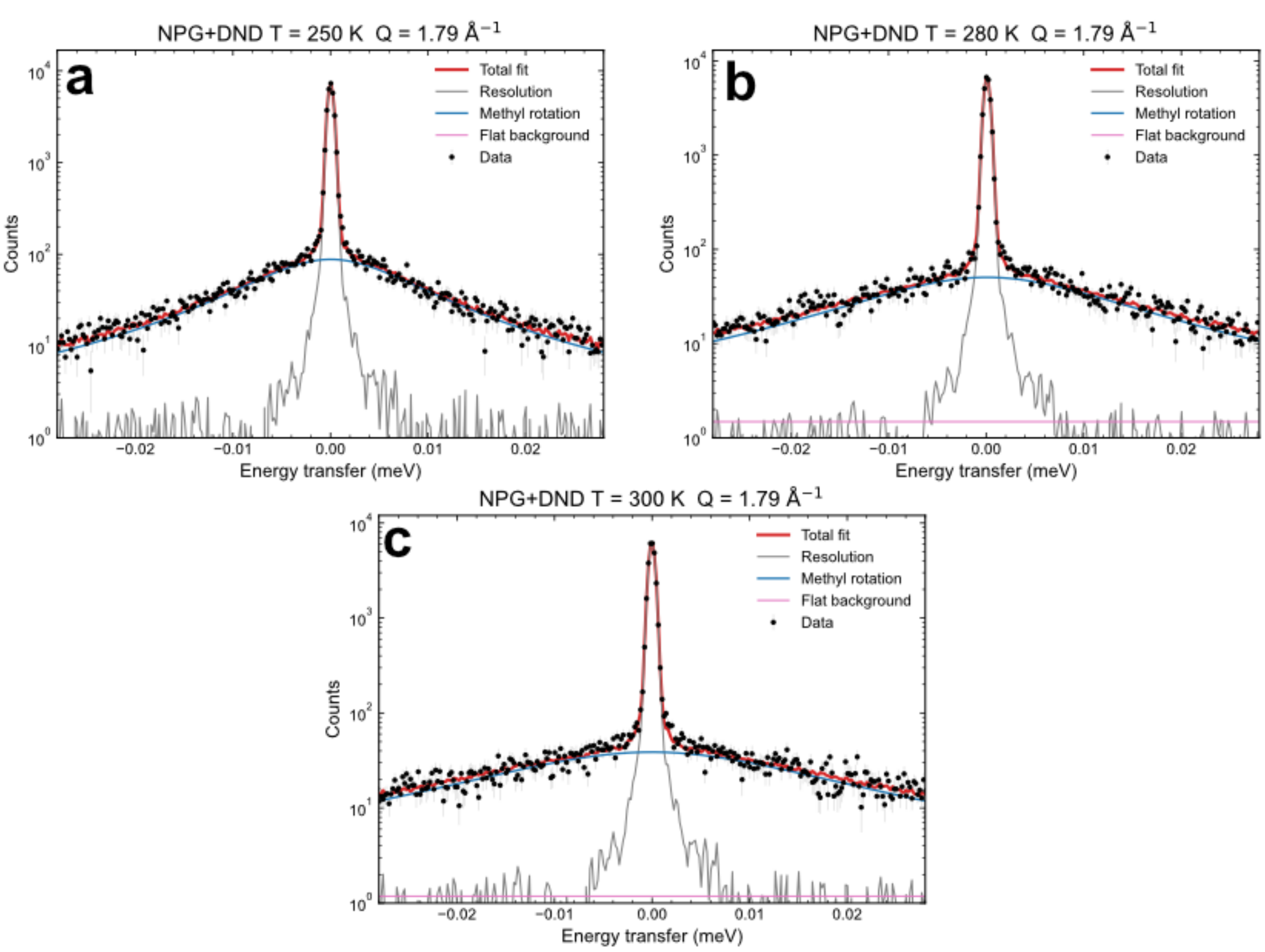

**Figure S8**: Representative QENS fits ($Q$ = 1.79 Å$^{-1}$) for NPG+DND in the OC phase at (a) 250 K, (b) 280 K and (c) 300 K. Only a single Lorentzian is included in the model at these temperatures, accounting for the methyl rotation. The flat background term accounts for multiple-scattering and other inelastic scattering contributions.

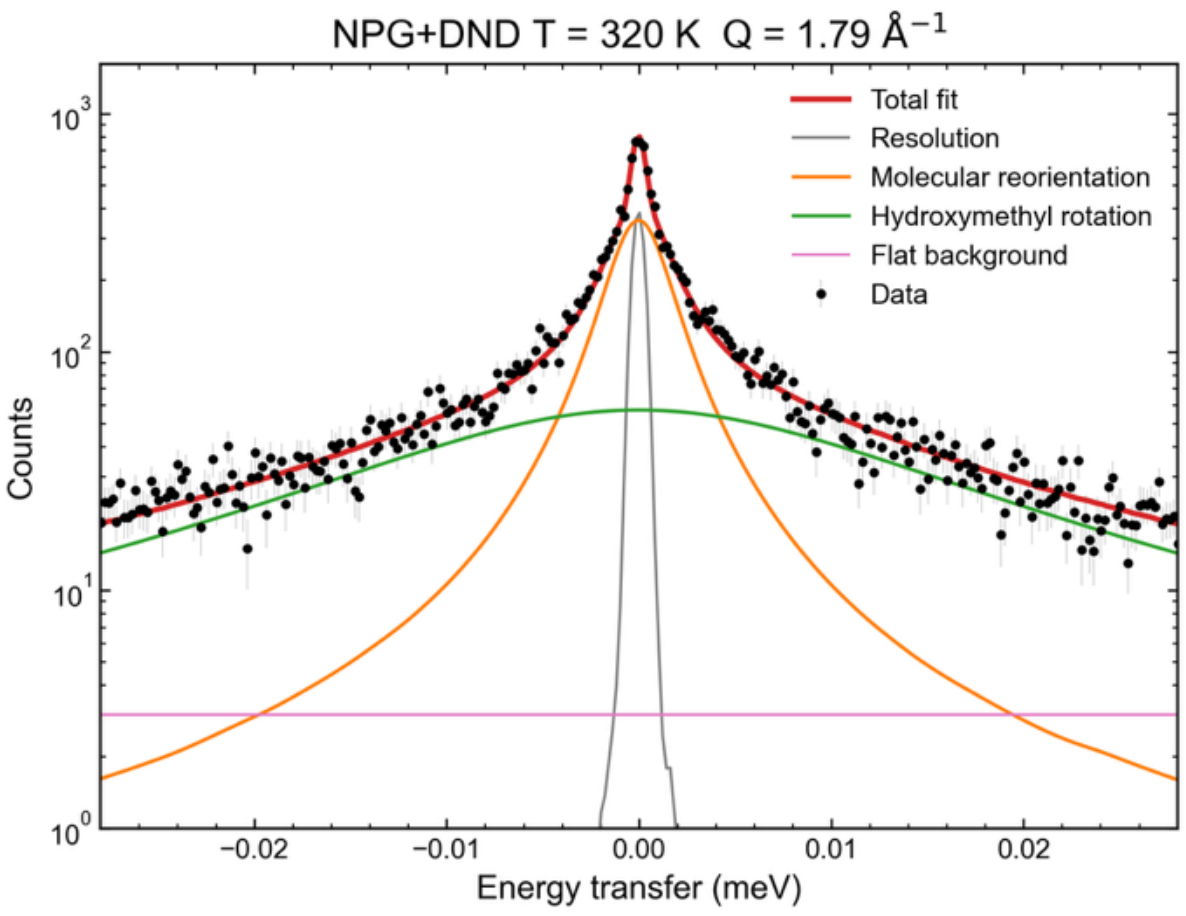

**Figure S9**: Representative QENS fit ($Q$ = 1.79 Å$^{-1}$) for NPG+DND in the PC phase. Two Lorentzians are included in the model at these temperatures, accounting for the hydroxymethyl rotation and molecular reorientation. Methyl rotation is present in the PC phase but is not observed as part of the QENS signal as a Lorentzian component, since this mode is outside the dynamic range of the instrument and is therefore included in the flat background.

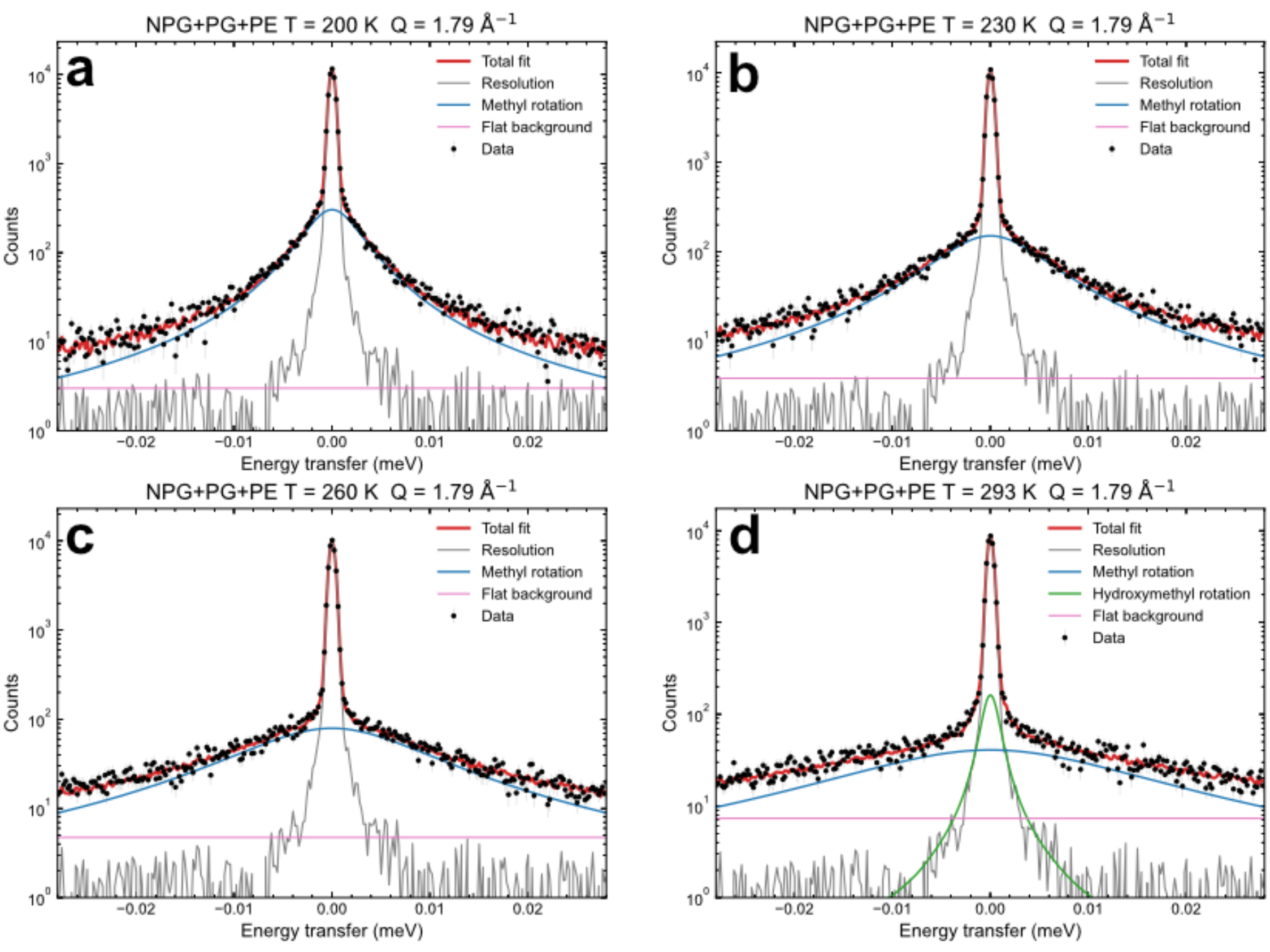

**Figure S10**: Representative QENS fits ($Q$ = 1.79 Å$^{-1}$) for NPG+PG+PE in the OC phase at (a) 200 K, (b) 230 K and (c) 260 K. Only a single Lorentzian is included in the model at these temperatures, accounting for the methyl rotation. (d) A separate QENS fit for $T$ = 293 K, in the hysteresis region, not included in the global fit. Here, two Lorentzians are required to adequately fit the data, one consistent with OC phase methyl rotation and the other likely accounts for both hydroxymethyl rotation and molecular reorientation. The fit in (d) demonstrates that these modes are slowly liberated before the bulk phase transition occurs in NPG+PG+PE. The flat background term accounts for multiple-scattering and other inelastic scattering contributions.

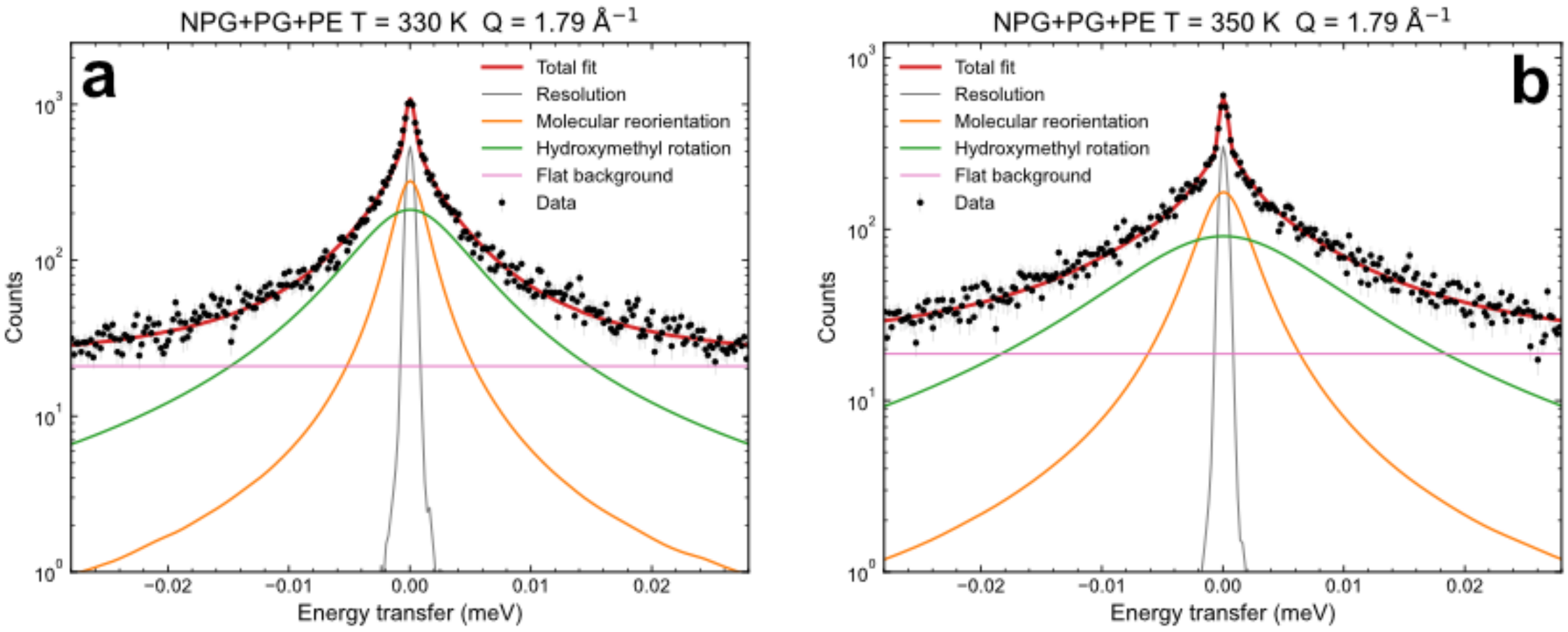

**Figure S11**: Representative QENS fits ($Q$ = 1.79 Å$^{-1}$) for NPG+PG+PE in the PC phase at (a) 330 K, (b) 350 K. Two Lorentzians are included in the model at these temperatures, accounting for the hydroxymethyl rotation and molecular reorientation. Methyl rotation is present in the PC phase but is not observed as part of the QENS signal as a Lorentzian component, since this mode is outside the dynamic range of the instrument and is therefore included in the flat background.

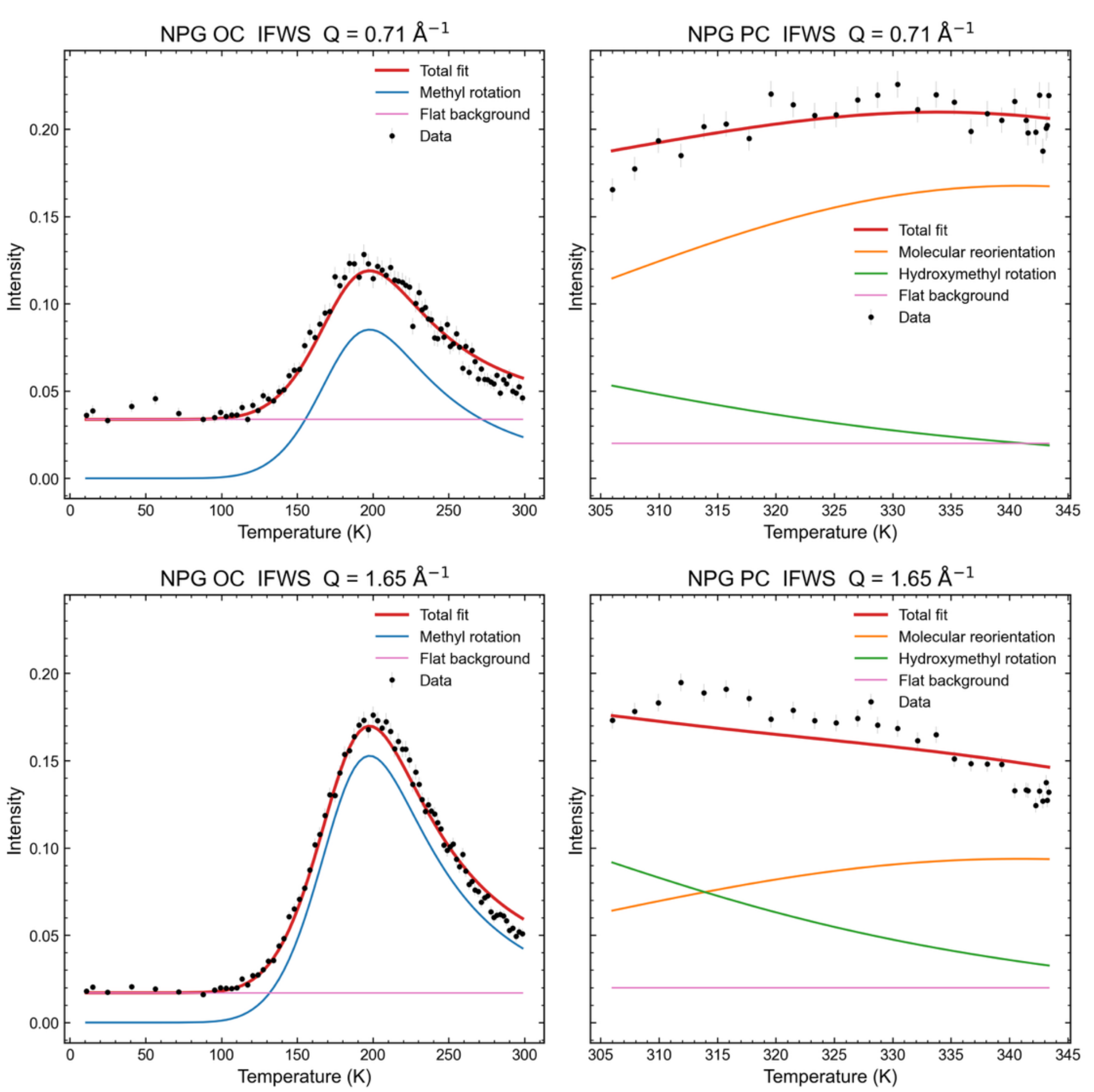

**Figure S12**: Representative IFWS fits ($Q$ = 0.71 Å$^{-1}$ and $Q$ = 1.65 Å$^{-1}$) for NPG. Left panels show the fits for the OC phase ($T<T_0$) and right panels show the fits for the PC phase ($T>T_0$). As with the QENS fits, one mode is modelled in the OC phase (methyl rotation) whereas two modes are modelled in the PC phase (hydroxymethyl rotation and molecular reorientation). These IFWS data were globally fitted with the corresponding QENS data shown in Figures S4-S5, as described in the main manuscript text.

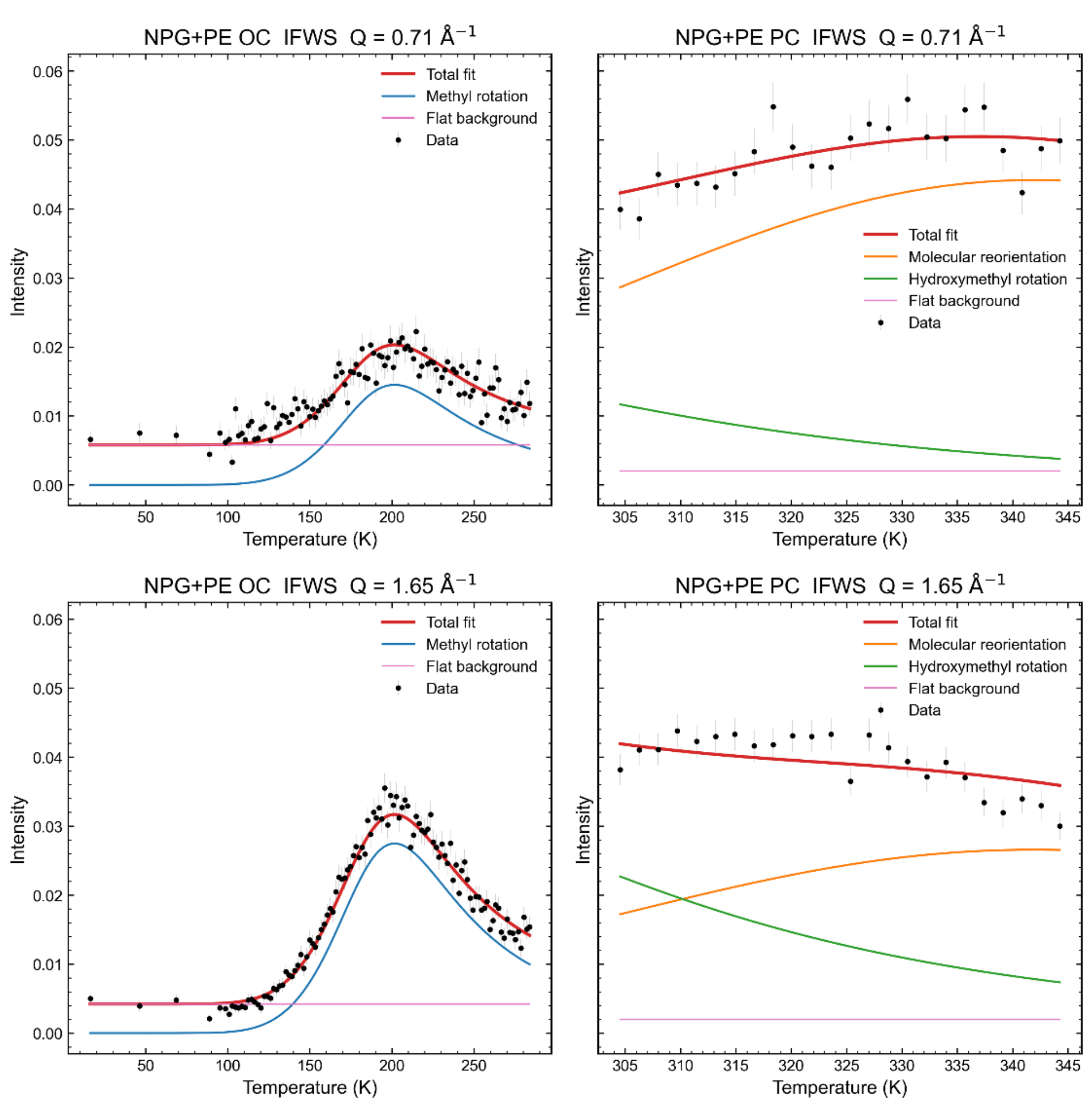

**Figure S13**: Representative IFWS fits ($Q$ = 0.71 Å$^{-1}$ and $Q$ = 1.65 Å$^{-1}$) for NPG+PE. Left panels show the fits for the OC phase ($T$<$T_0$) and right panels show the fits for the PC phase ($T$ >$T_0$). As with the QENS fits, one mode is modelled in the OC phase (methyl rotation) whereas two modes are modelled in the PC phase (hydroxymethyl rotation and molecular reorientation). These IFWS data were globally fitted with the corresponding QENS data shown in Figures S6-S7, as described in the main manuscript text.

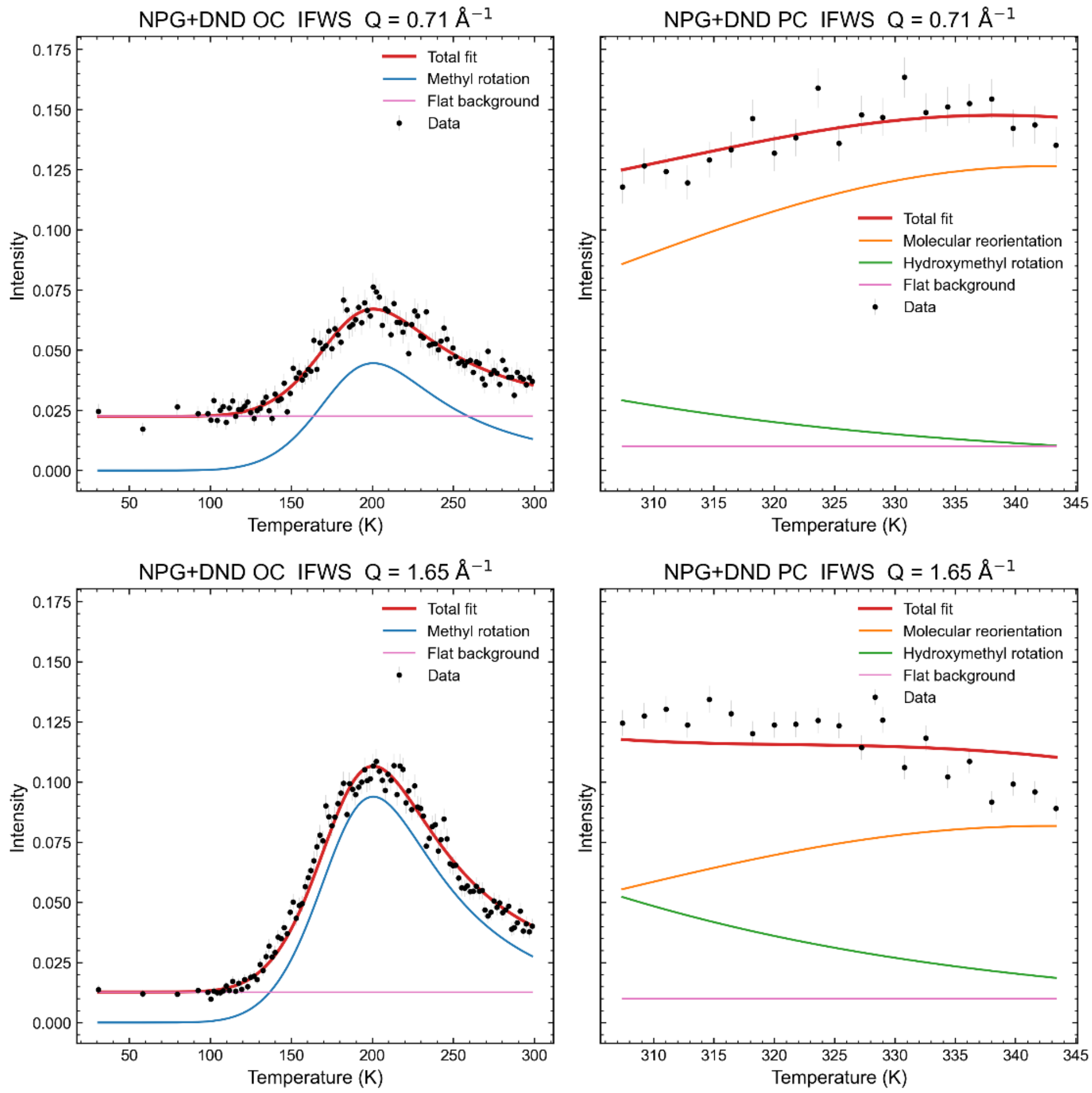

**Figure S14**: Representative IFWS fits ($Q$ = 0.71 Å$^{-1}$ and $Q$ = 1.65 Å$^{-1}$) for NPG+DND. Left panels show the fits for the OC phase ($T$<$T_0$) and right panels show the fits for the PC phase ($T$ >$T_0$). As with the QENS fits, one mode is modelled in the OC phase (methyl rotation) whereas two modes are modelled in the PC phase (hydroxymethyl rotation and molecular reorientation). These IFWS data were globally fitted with the corresponding QENS data shown in Figures S8-S9, as described in the main manuscript text.

**Figure S15**: Representative IFWS fits ($Q$ = 0.71 Å$^{-1}$ and $Q$ = 1.65 Å$^{-1}$) for NPG+PG+PE. Left panels show the fits for the OC phase ($T<T_0$) and right panels show the fits for the PC phase ($T>T_0$). As with the QENS fits, one mode is modelled in the OC phase (methyl rotation) whereas two modes are modelled in the PC phase (hydroxymethyl rotation and molecular reorientation). These IFWS data were globally fitted with the corresponding QENS data shown in Figures S10-S11, as described in the main manuscript text.

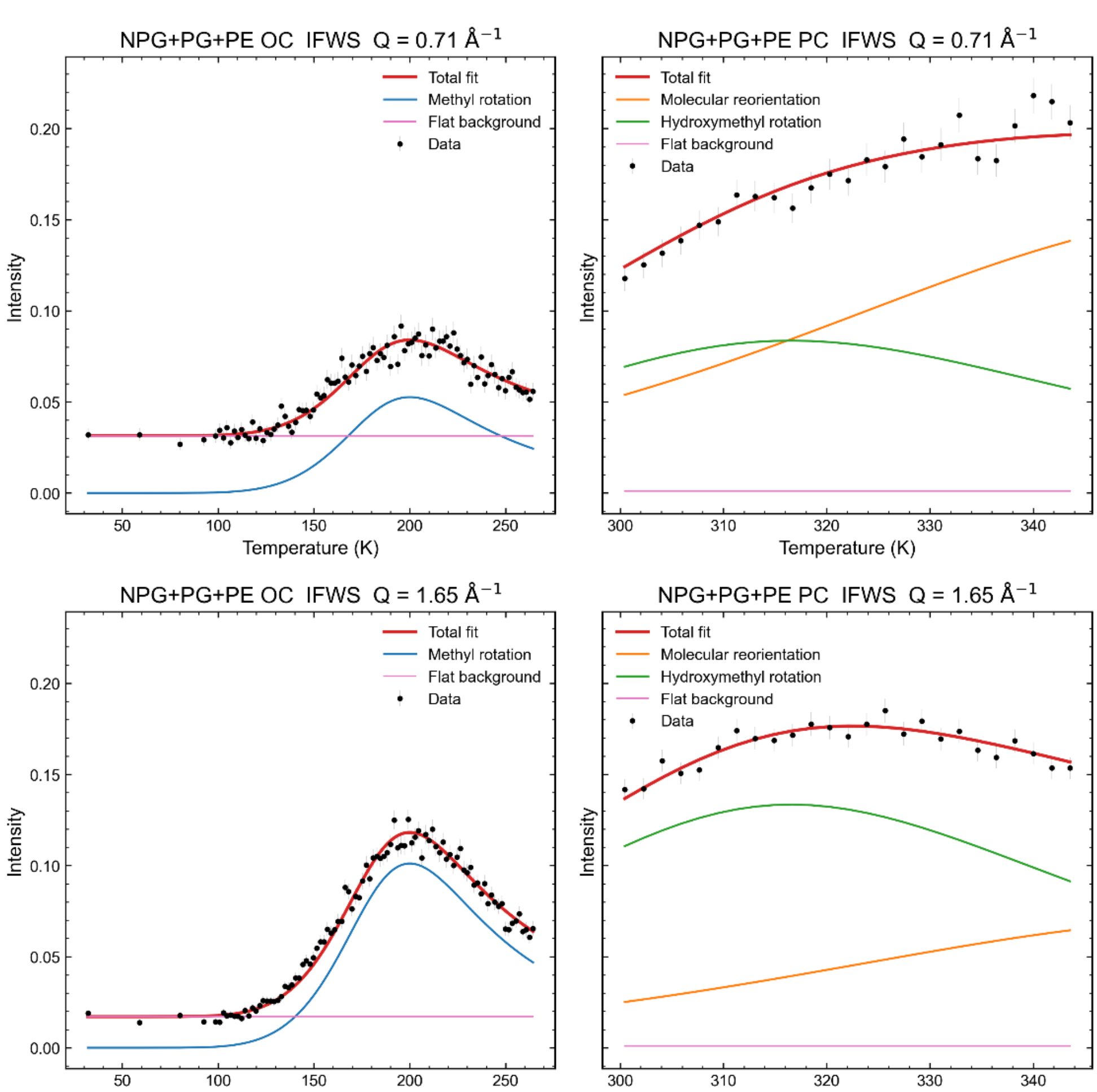

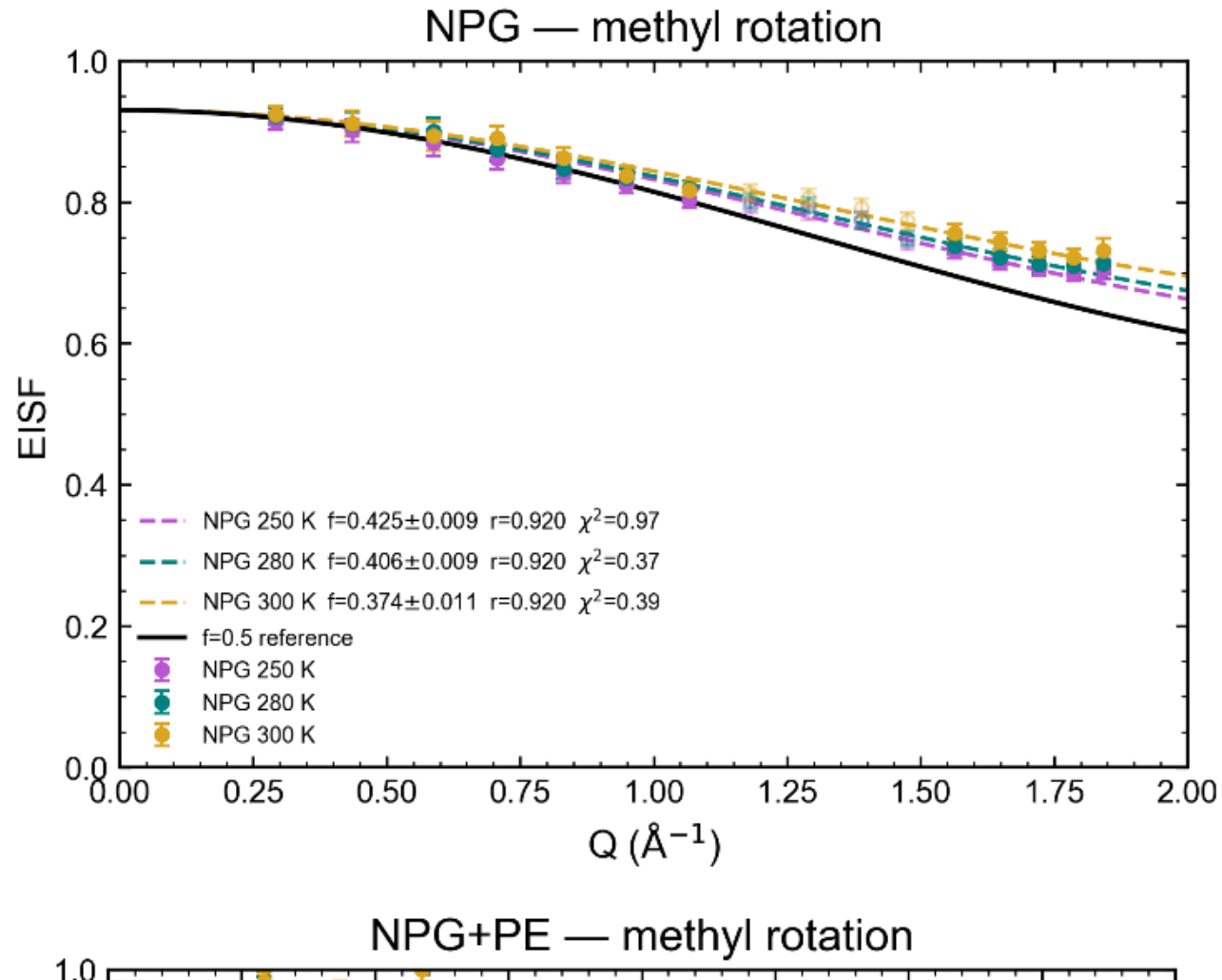

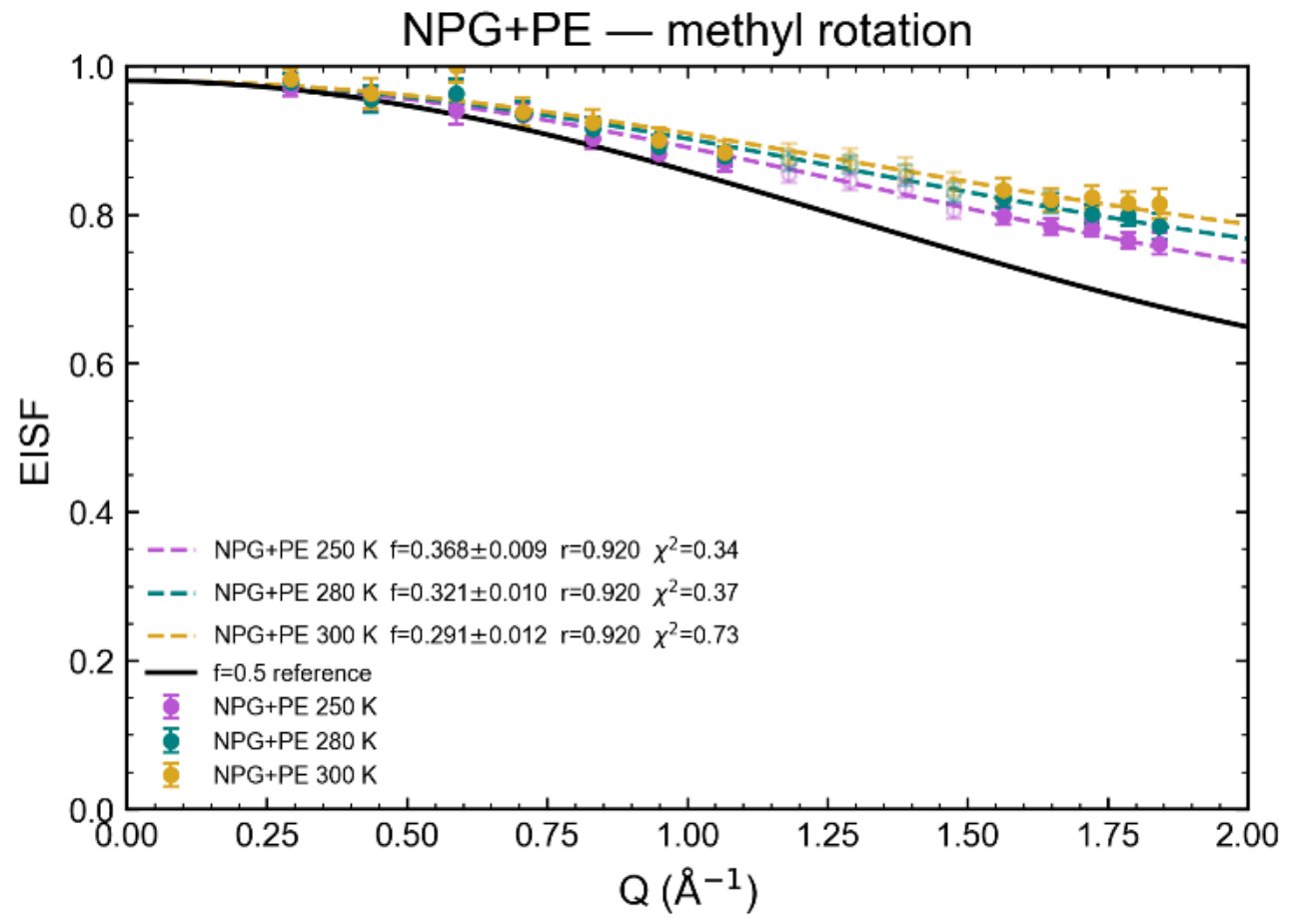

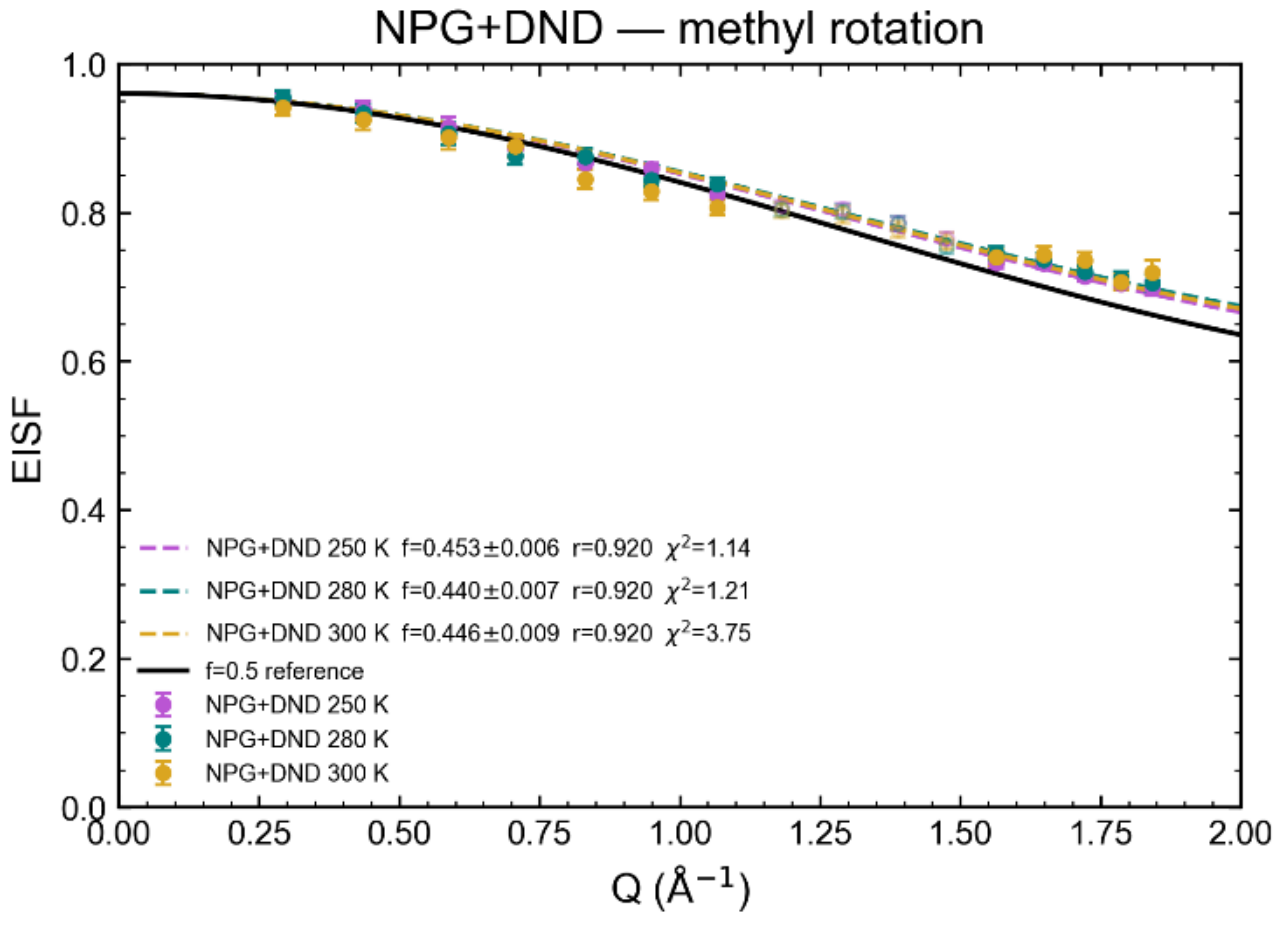

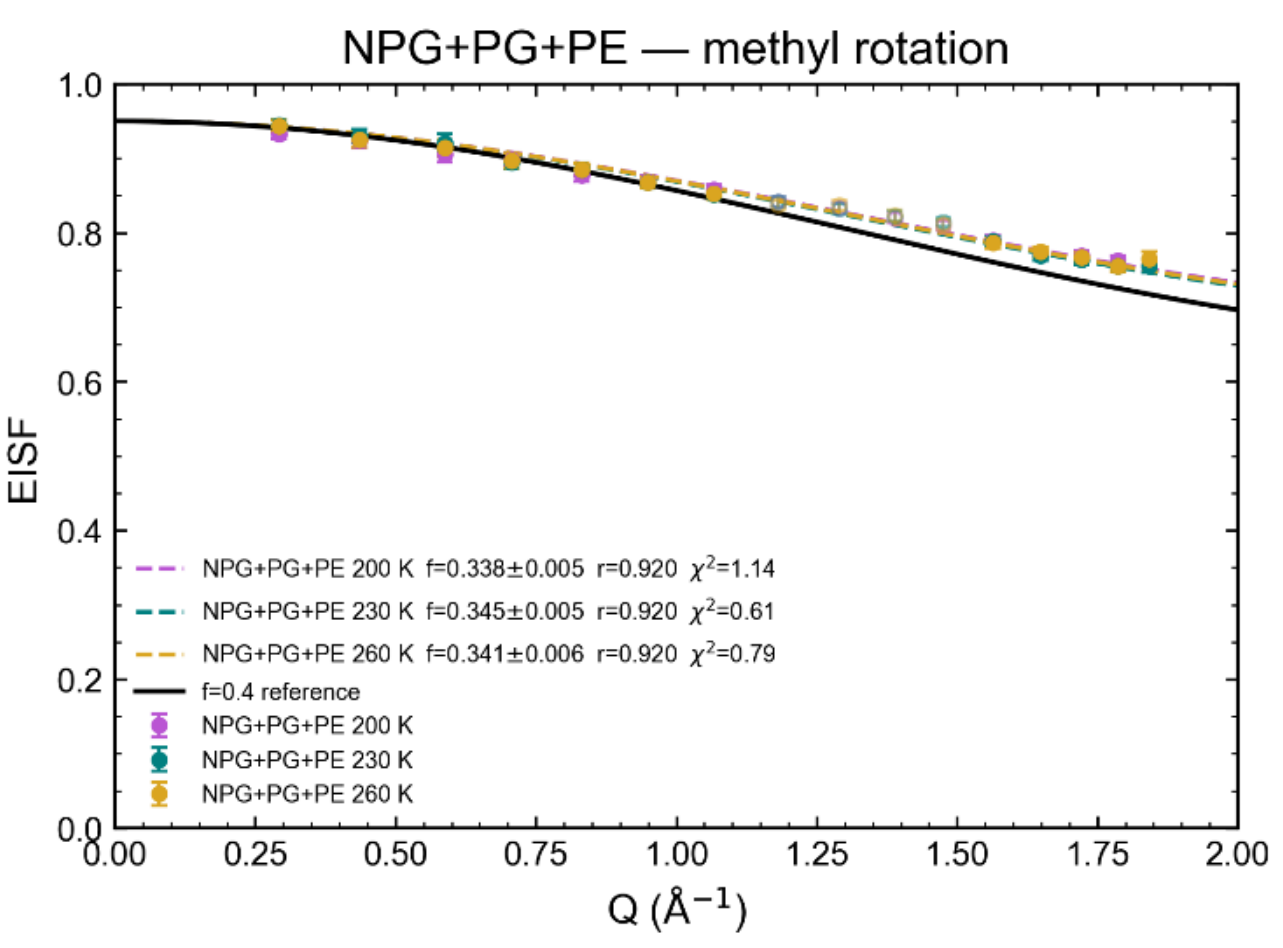

**Figure S16**: EISF plots for OC methyl rotation at the various measured temperatures in NPG (0.93), NPG+PE (0.98), NPG+DND (0.96) and NPG+PG+PE (0.95). A linear scaling factor was required to provide reasonable fits to the data to account for the fact that the EISF does not approach unit at zero *Q*, given in brackets. This is likely due to some multiple scattering effects that cause some non-zero QENS signal even at low *Q*. The solid black line indicates a reference for the maximum mobile fraction of hydrogen scatterers associated with methyl groups, since the hydroxymethyl hydrogens are locked in hydrogen bonds in the OC phase and are therefore static on the timescale of the instrument. Empty data points between 1.1 Å$^{-1}$ < *Q* < 1.5 Å$^{-1}$ were masked and not included in the fits, due to the presence of NPG Bragg peaks in this *Q* range.

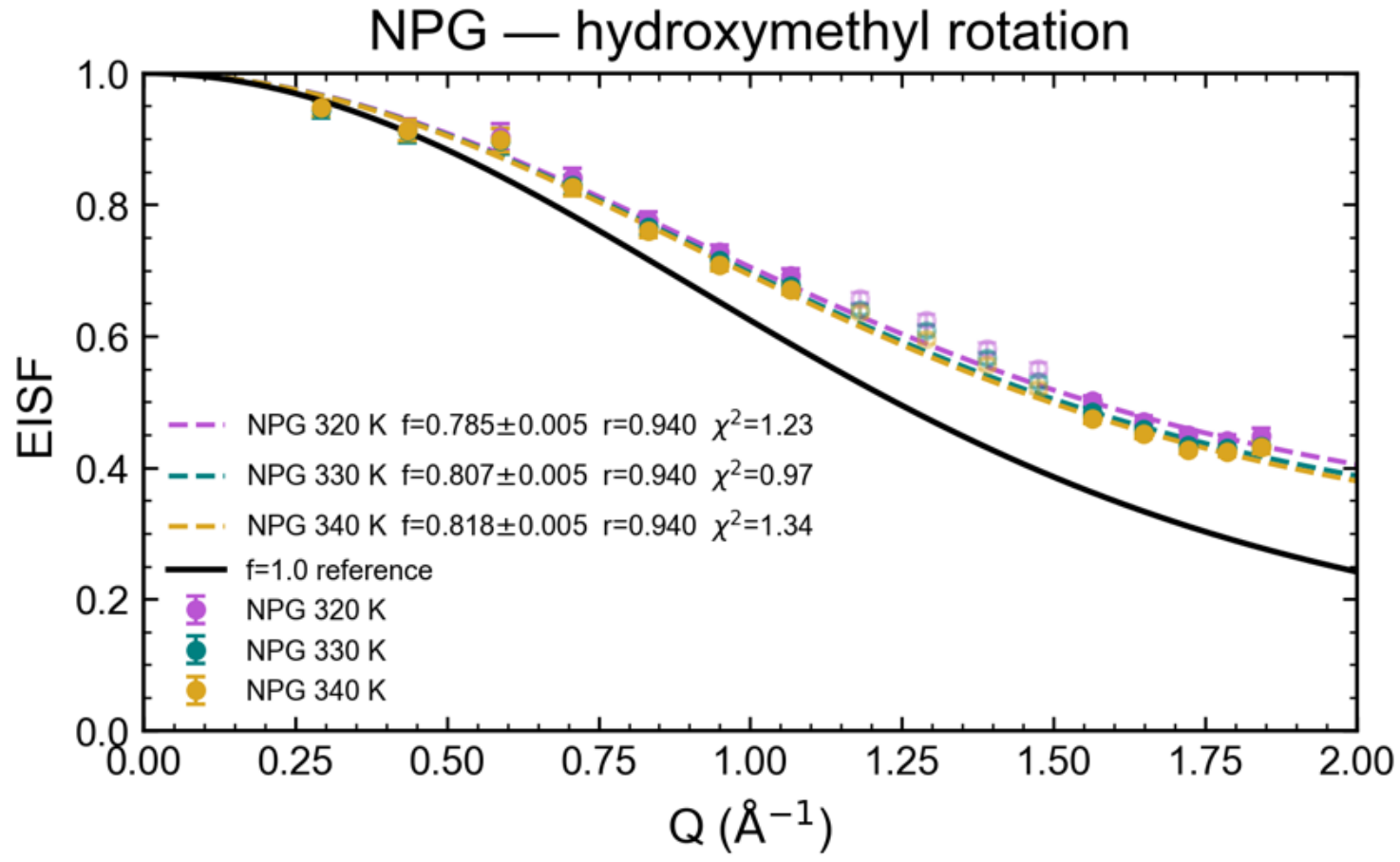

**Figure S17**: EISF plots for PC hydroxymethyl rotation at the various measured temperatures in NPG, NPG+PE, NPG+DND and NPG+PG+PE. The solid black line indicates a reference for the maximum mobile fraction of hydrogen scatterers associated with this motion. All samples show a similar insensitivity of the mobile fraction with temperature, except NPG+PE, which displays a more significant temperature variation. Empty data points between 1.1 Å$^{-1}$ < $Q$ < 1.5 Å$^{-1}$ were masked and not included in the fits, due to the presence of NPG Bragg peaks in this $Q$ range.

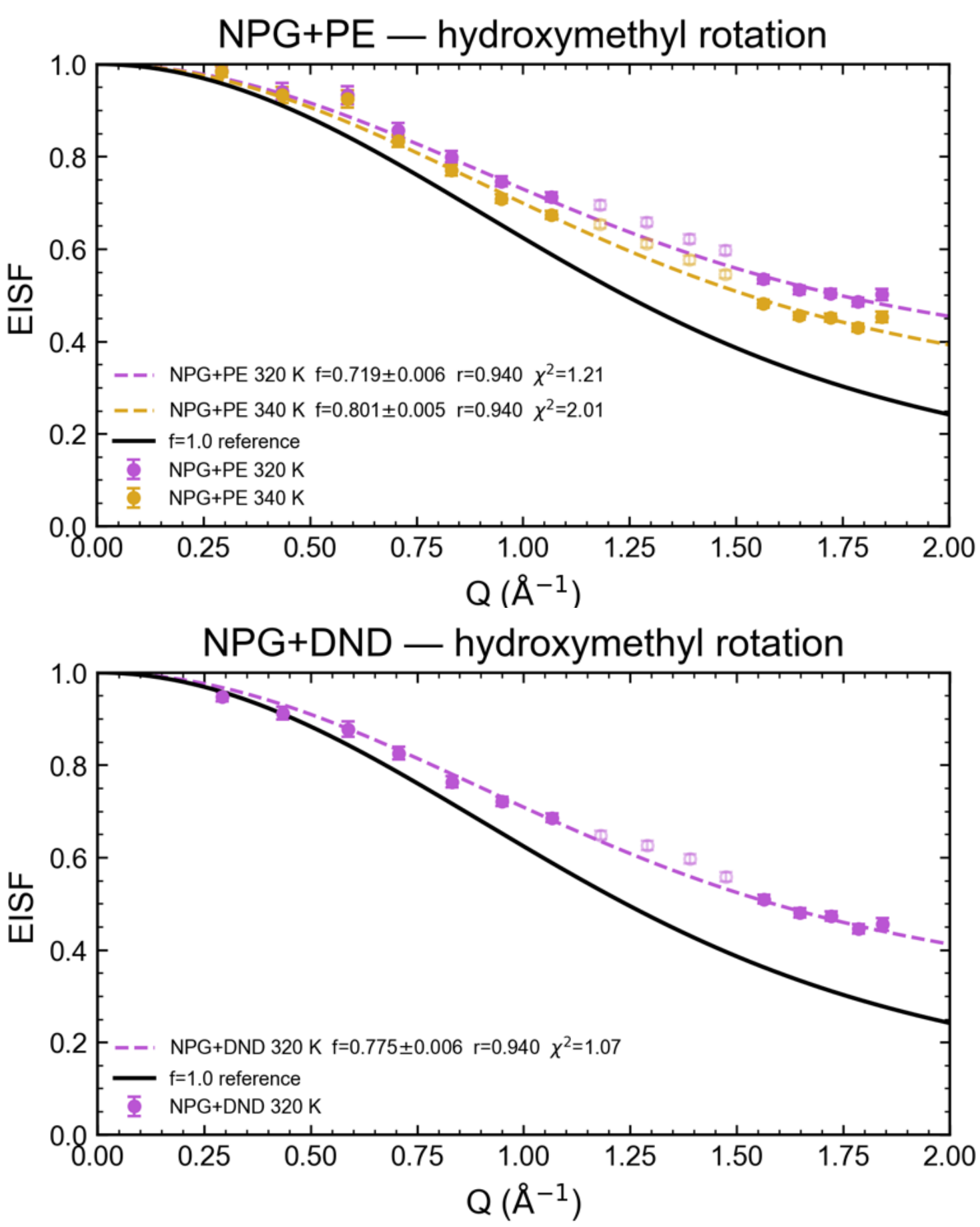

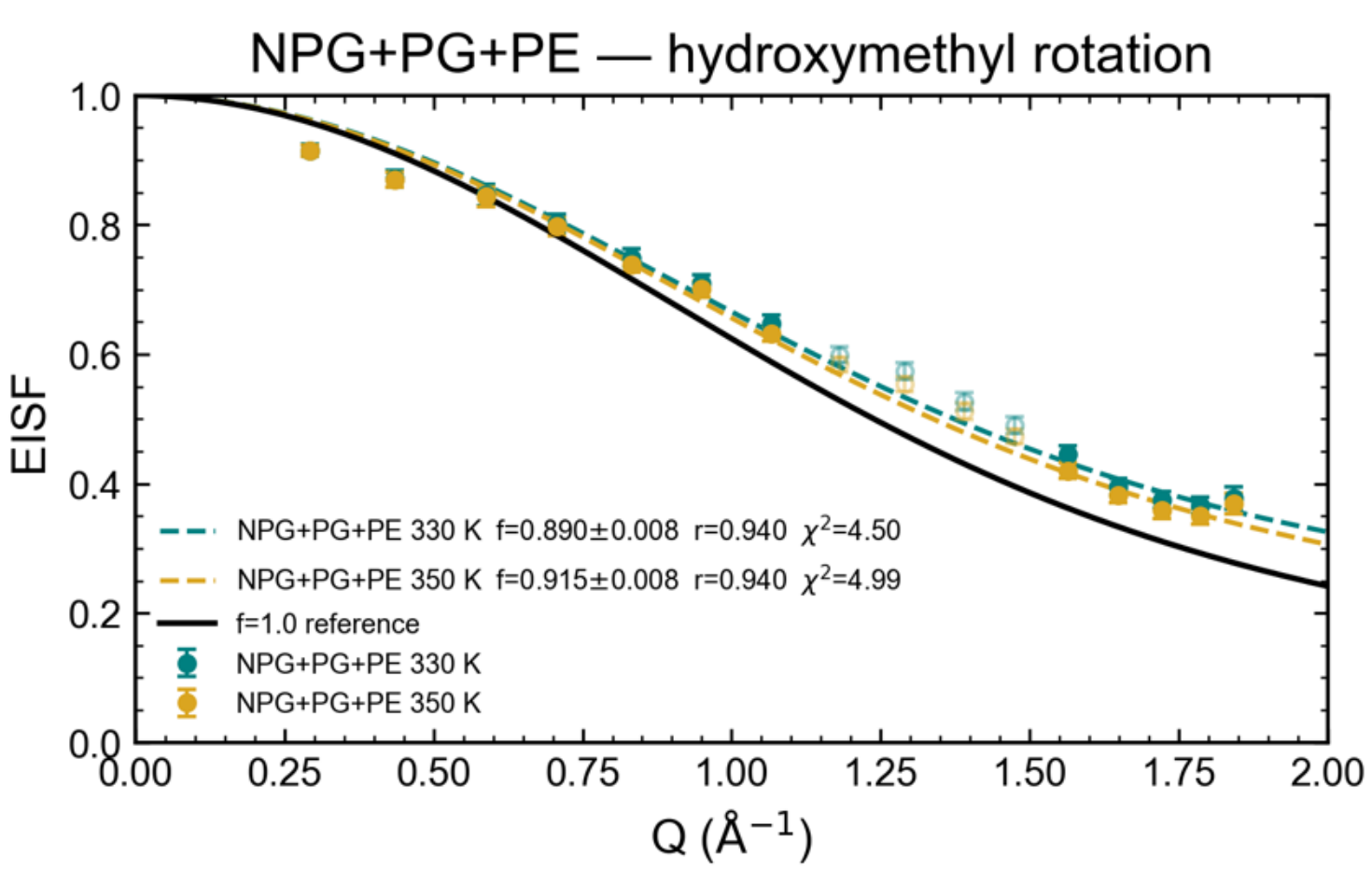

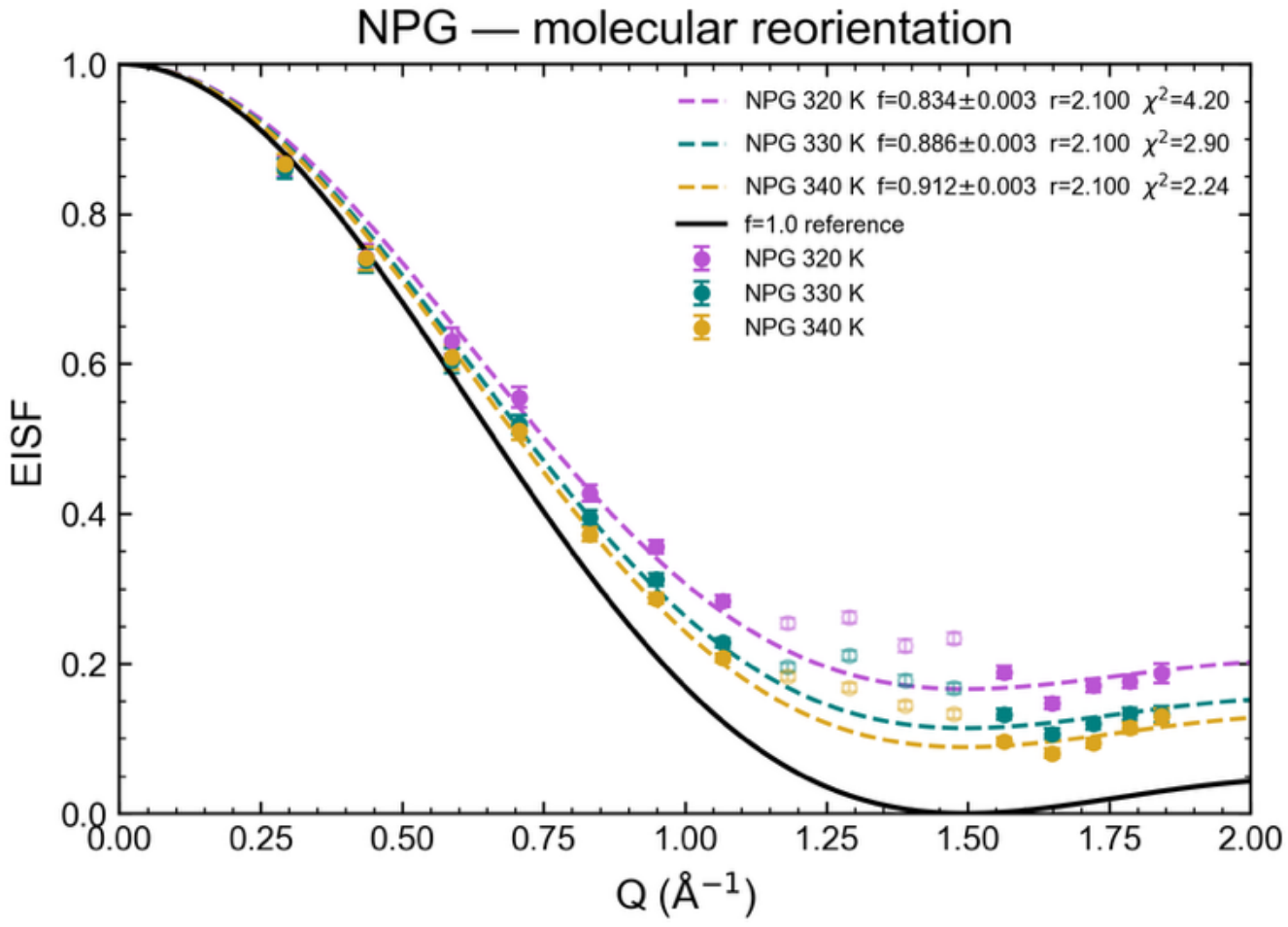

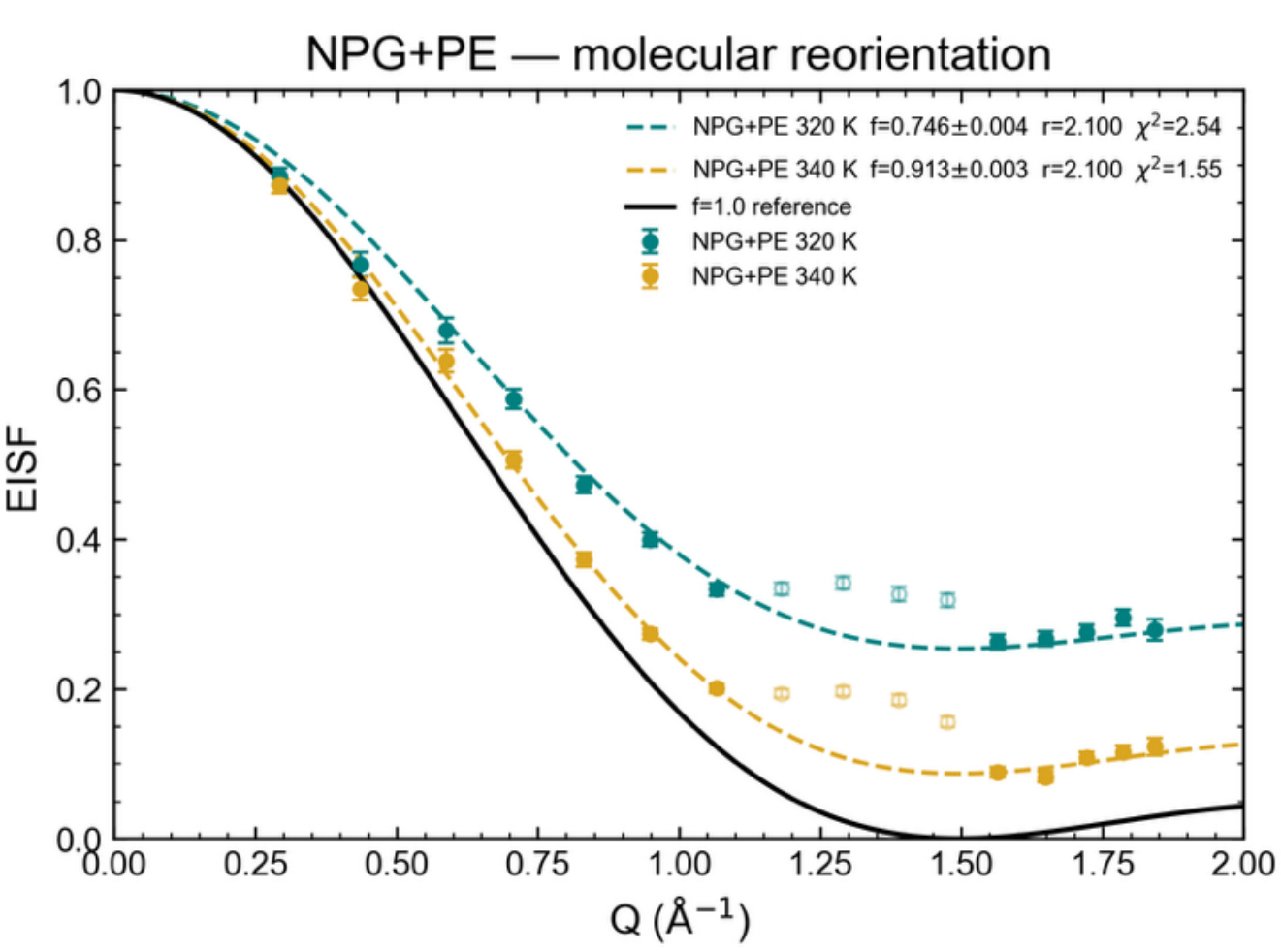

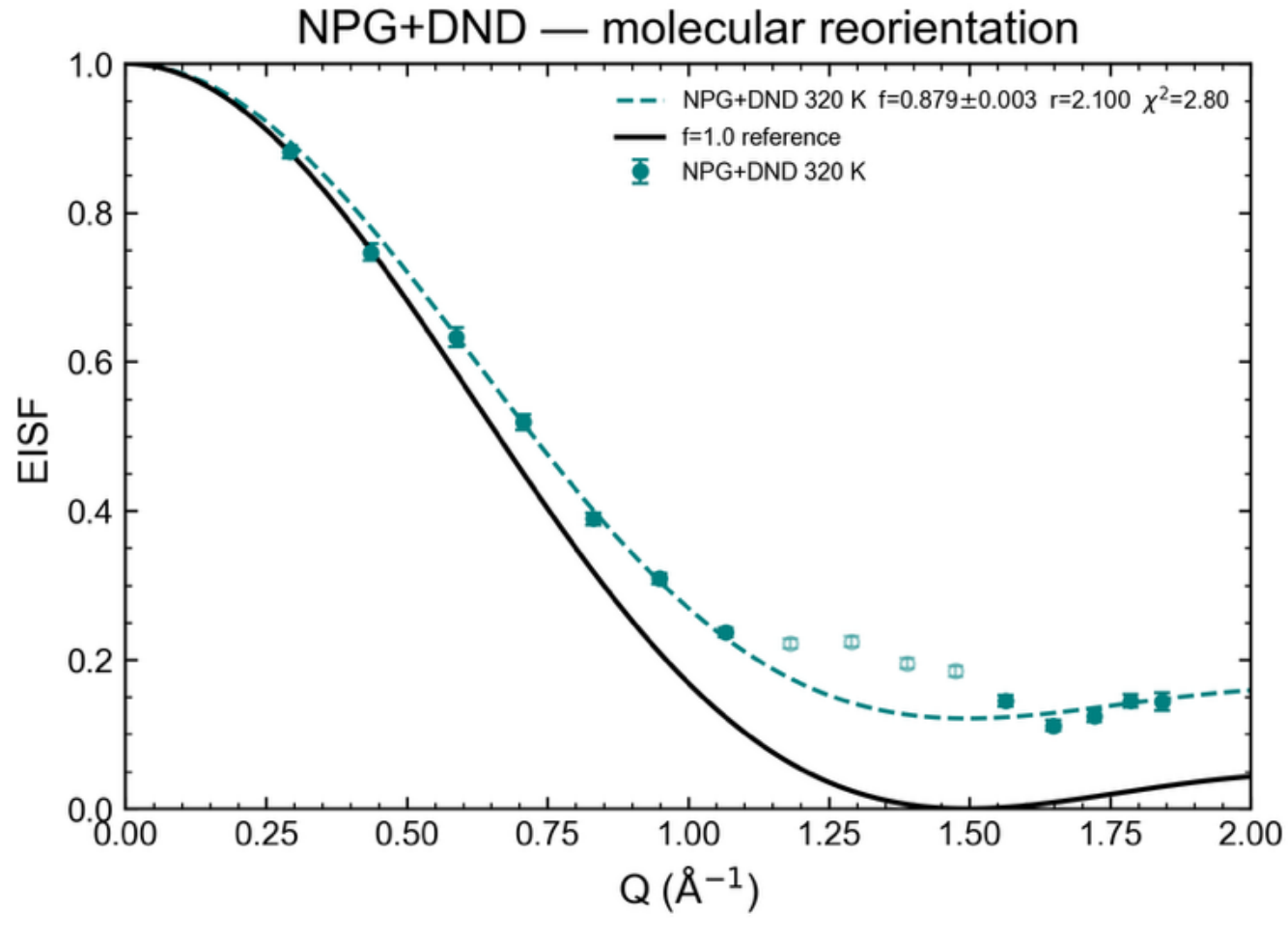

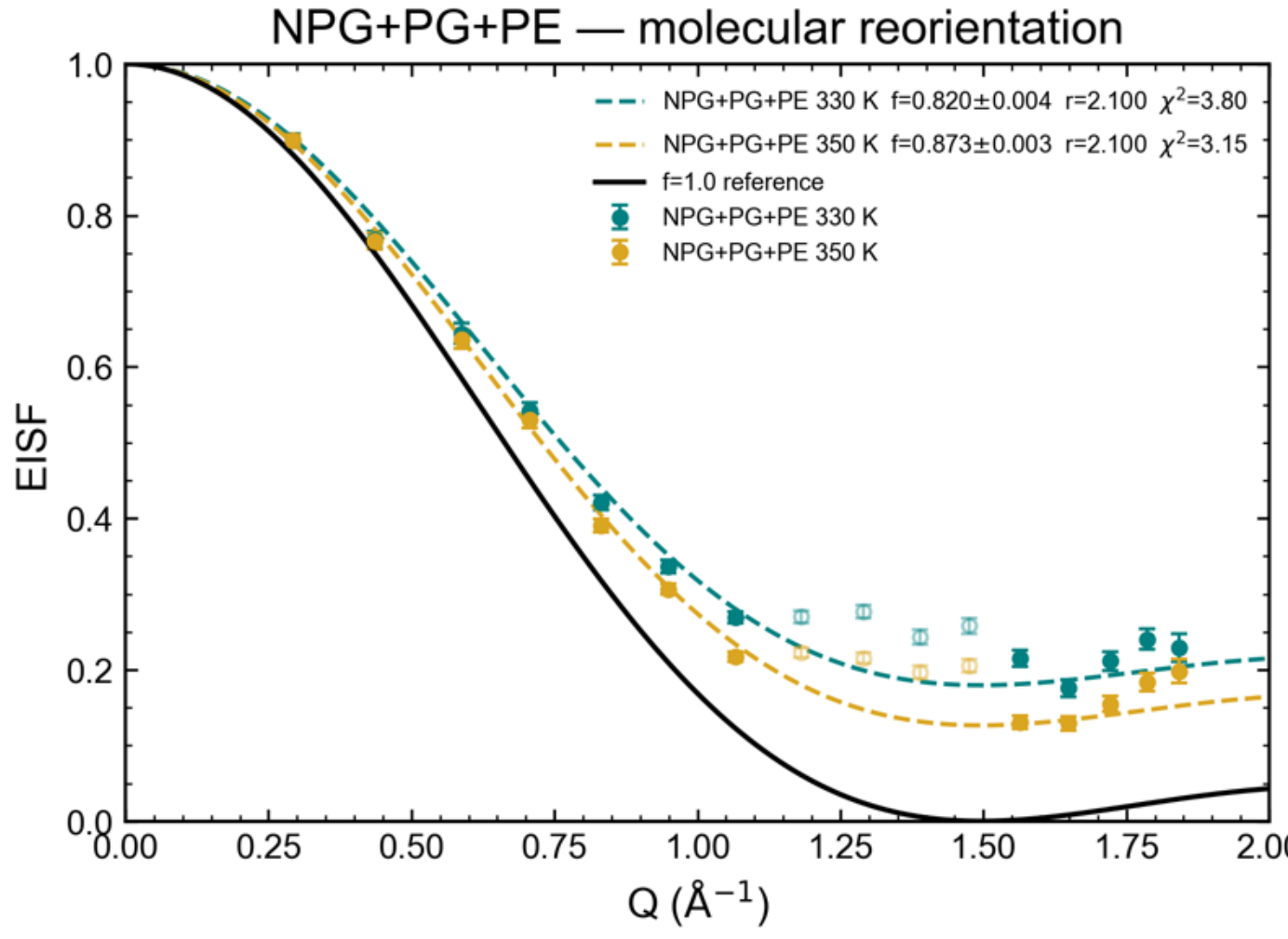

**Figure S18**: EISF plots for PC molecular reorientation at the various measured temperatures in NPG, NPG+PE, NPG+DND and NPG+PG+PE. The solid black line indicates a reference for the maximum mobile fraction of hydrogen scatterers associated with this motion. All samples display relatively strong variation in the mobile fraction of scatterers with increasing temperature. This suggests that the mobility of the molecular tumbling mode is not fully liberated immediately after the OC-PC phase transition. Empty data points between 1.1 Å$^{-1}$ < $Q$ < 1.5 Å$^{-1}$ were masked and not included in the fits, due to the presence of NPG Bragg peaks in this $Q$ range.

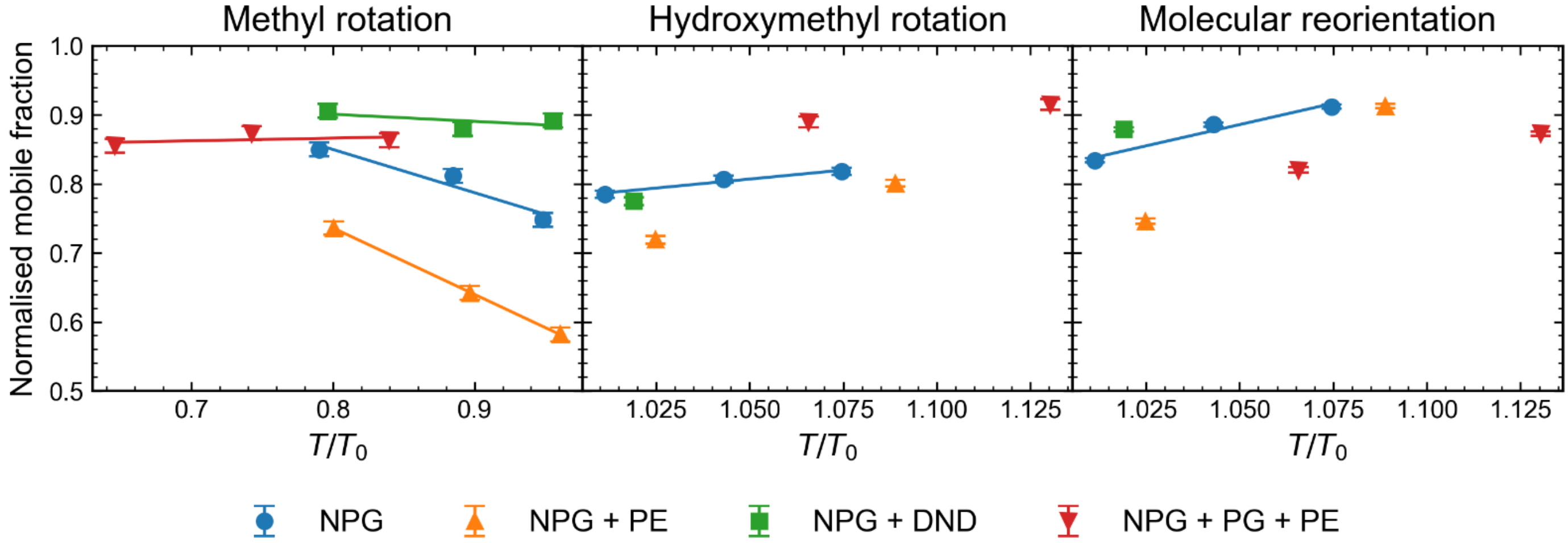

**Figure S19**: A comparison of the mobile fraction of scatters as a function of reduced temperature, where $T_0$ is the equilibrium phase transition temperature extracted from the EFWS fits given in Fig. S3 for each sample, summarising the results shown in Figures S16-S18. Solid lines indicate linear best fits to datasets with more than two data points. For methyl rotation, the mobile fraction was normalised to the ratio of methyl groups able to rotate in the OC phase, 0.500 for NPG, NPG+PE, NPG+DND and 0.395 for NPG+PG+PE, since the remaining hydrogens are locked in hydrogen bonds in the OC phase and are static on the timescale of the instrument. The fact that the mobile fraction of methyl rotation appears to decrease with increasing temperature is unexpected behaviour that we cannot definitively explain here.